\documentclass[atoms,article,submit,moreauthors, pdftex,10pt,a4paper]{mdpi} 

\usepackage{epsfig}

\usepackage{color}

\usepackage{graphicx}
\usepackage{amsmath}
\usepackage{amssymb}
\usepackage{epstopdf}
\usepackage{dcolumn}
\usepackage{bm}
\usepackage{verbatim}
\usepackage{amsfonts}
\usepackage{cancel}
\usepackage[utf8]{inputenc}
\usepackage{graphicx}
\usepackage{setspace}
\usepackage{tabularx}
\usepackage{colortbl}
\usepackage[export]{adjustbox}

\newcommand{\acs}{\alpha_{\textrm{Cs}}(0)}

\newcommand{\ak}{\alpha_{\textrm{K}}(0)}
\newcommand{\arcs}{\alpha_{r,\textrm{Cs}}(0)}
\newcommand{\arrb}{\alpha_{r,\textrm{Rb}}(0)}
\newcommand{\ark}{\alpha_{r,\textrm{K}}(0)}
\newcommand{\arna}{\alpha_{r,\textrm{Na}}(0)}
\newcommand{\arli}{\alpha_{r,\textrm{Li}}(0)}
\newcommand{\ccs}{C_{6,\textrm{Cs}}}
\newcommand{\crb}{C_{6,\textrm{Rb}}}
\newcommand{\ck}{C_{6,\textrm{K}}}
\newcommand{\etal}{et \emph{al.}~}
\newcommand{\etalnospace}{ et \emph{al.}}
\newcommand{\beq}{\begin{equation}}
\newcommand{\eeq}{\end{equation}}
\newcommand{\figref}[1]{Fig.~\ref{#1}}
\newcommand{\eqnref}[1]{Eqn.~\eqref{#1}}

\newcommand{\abinitio}{\textit{ab initio }}

\firstpage{1} 
\articlenumber{x}
\doinum{10.3390/------}
\pubvolume{xx}
\pubyear{2016}
\copyrightyear{2016}
\externaleditor{Academic Editor: name}
\history{Received: date; Accepted: date; Published: date}

 \theoremstyle{mdpi}
 \newcounter{thm}
 \newcounter{ex}
 \newcounter{re}
 \theoremstyle{mdpidefinition}

\Title{Analysis of polarizability measurements made with atom interferometry}

\Author{Maxwell D. Gregoire $^{1}$, Nathan Brooks $^{1}$, Raisa Trubko$^{2}$, and Alexander D. Cronin $^{1,2}$*}
\AuthorNames{Maxwell D. Gregoire, Nathan Brooks, Raisa Trubko, and Alexander D. Cronin}

\address{%
$^{1}$ \quad Department of Physics, University of Arizona, Tucson, Arizona\\
$^{2}$ \quad College of Optical Sciences, University of Arizona, Tucson, Arizona}

\corres{Correspondence: cronin@physics.arizona.edu; Tel.: 520-465-8459}

\abstract{
We present revised measurements of the static electric dipole polarizabilities of K, Rb, and Cs based on atom interferometer experiments presented in 
[\textit{Phys. Rev. A} \textbf{2015}, 92, 052513]
but now re-analyzed with new calibrations for the magnitude and geometry of the applied electric field gradient. The resulting polarizability values did not change, but the uncertainties were significantly reduced.
Then we interpret several measurements of alkali metal atomic polarizabilities in terms of atomic oscillator strengths $f_{ik}$, Einstein coefficients $A_{ik}$, state lifetimes $\tau_{k}$, transition dipole matrix elements $D_{ik}$, line strengths $S_{ik}$, and van der Waals $C_6$ coefficients. Finally, we combine atom interferometer measurements of polarizabilities with independent measurements of lifetimes and $C_6$ values in order to quantify the residual contribution to polarizability due to all atomic transitions other than the principal $ns$-$np_J$ transitions for alkali metal atoms.  }


\begin{document}



\section{Introduction} \label{sec:intro}

Atomic and molecular interferometry \cite{Berman1997,Cronin2009} has become a precise method for measuring atomic properties such as static polarizabilities \cite{Gregoire2015,Ekstrom1995,Miffre2006,Berninger2007,Holmgren2010}, van der Waals interactions \cite{Perreault2005, Lepoutre2009, Lepoutre2011}, and tune-out wavelengths \cite{Holmgren2012,Leonard2015}.  Calculating these atomic and molecular properties \emph{ab initio}  is challenging because it requires modeling of quantum many-body systems with relativistic corrections.  For example, different methods for calculating polarizabilities yield results that vary by as much as  10\% for Cs
\cite{Mitroy2010, tang2009, sahoo2007, hamonou2007, deiglmayr2008, Johnson2008, Reinsch1976, Tang1976,Maeder1979,  Christiansen1982, Fuentealba1982, Muller1984, kundu1986,  Kello1993, Fuentealba1993, Marinescu1994, Dolg1996, Patil1997, Lim1999,  Safronova1999, Derevianko1999, Magnier2002, Mitroy2003, Safronova2004, Lim2005, Arora2007, Iskrenova-Tchoukova2007, Safronova2008, Safronova2011}.  For molecules the challenges are even greater.  Furthermore, determining the uncertainty for an \emph{ab initio} calculation can be difficult.  Polarizability measurements made with matter wave interferometry, therefore, have been used to assess which calculation methods are most valid.  Testing these calculations is important because similar methods are used to predict atomic scattering cross sections, Feshbach resonances, photoassociation rates,  atom-surface van-der Waals $C_3$ coefficients, atomic parity-violating amplitudes, and atomic clock shifts due to thermal radiation or collisions.

\begin{figure}
\centering
\includegraphics[width=\linewidth*3/4,keepaspectratio]{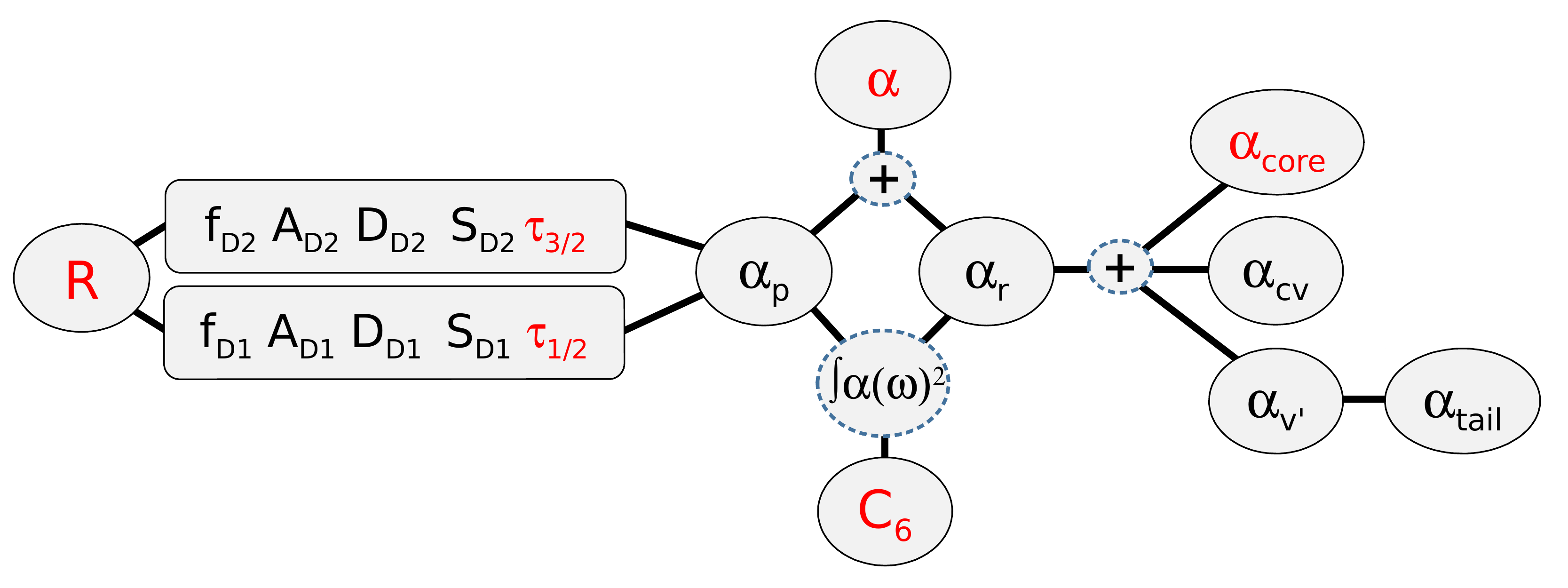}
\caption{\label{fig:ideachart} An idea chart showing connections between various quantities defined in Equations (\ref{eq:alphaf})-(\ref{eq:Rdef}) that we relate to static polarizability $\alpha(0)$ for alkali metal atoms.  Quantities in red have been directly measured. }
\end{figure}

In this manuscript, we first present revised uncertainties on our most recent K, Rb, and Cs static polarizability measurements \cite{Gregoire2015} in Section \ref{sec:revPolUnc}.  We then show how to use polarizability measurements for alkali metal atoms \cite{Ekstrom1995, Miffre2006, Holmgren2010, Gregoire2015} as input for semi-empirical calculations of atomic properties such as oscillator strengths, Einstein $A$ coefficients, state lifetimes, transition matrix elements, and line strengths, as we discuss in Section \ref{sec:fADS}.  We use polarizability measurements to predict van der Waals $C_6$ coefficients in Section \ref{sec:derivingC6}.
To support this analysis, throughout Section \ref{sec:pol2other} we use theoretical values for so-called \emph{residual polarizabilities} of alkali metal atoms, i.e.~the contributions to polarizabilities that come higher-energy excitations associated with the inner-shell (core) electrons and highly-excited states of the valence electrons.   The idea-chart in \figref{fig:ideachart} shows connections between the residual polarizability ($\alpha_r$) and several quantities related via Eqns.~(\ref{eq:alphaf})-(\ref{eq:Rdef}) that we use to interpret polarizabilities.

Then in Section \ref{sec:testing_ar}  we demonstrate an all-experimental method for measuring residual polarizabilities.  We do this by using polarizability measurements in combination with independent measurements of lifetimes and van der Waals $C_6$ coefficients.   This serves as a cross-check for some assumptions used in Section \ref{sec:pol2other} that are also used for analysis of atomic parity violation and atomic clocks.   Section \ref{sec:testing_ar}  highlights how atom interferometry measurements shown in Table \ref{tab:a0measurements} are sufficiently precise to directly measure the static residual polarizability,  $\alpha_r(0)$,  for each of the alkali-metal  atoms, Li, Na, K, Rb, and Cs.

\section{Revised uncertainties on recent polarizability measurements} \label{sec:revPolUnc}

\begin{figure}
\centering
\includegraphics[width=\linewidth,keepaspectratio]{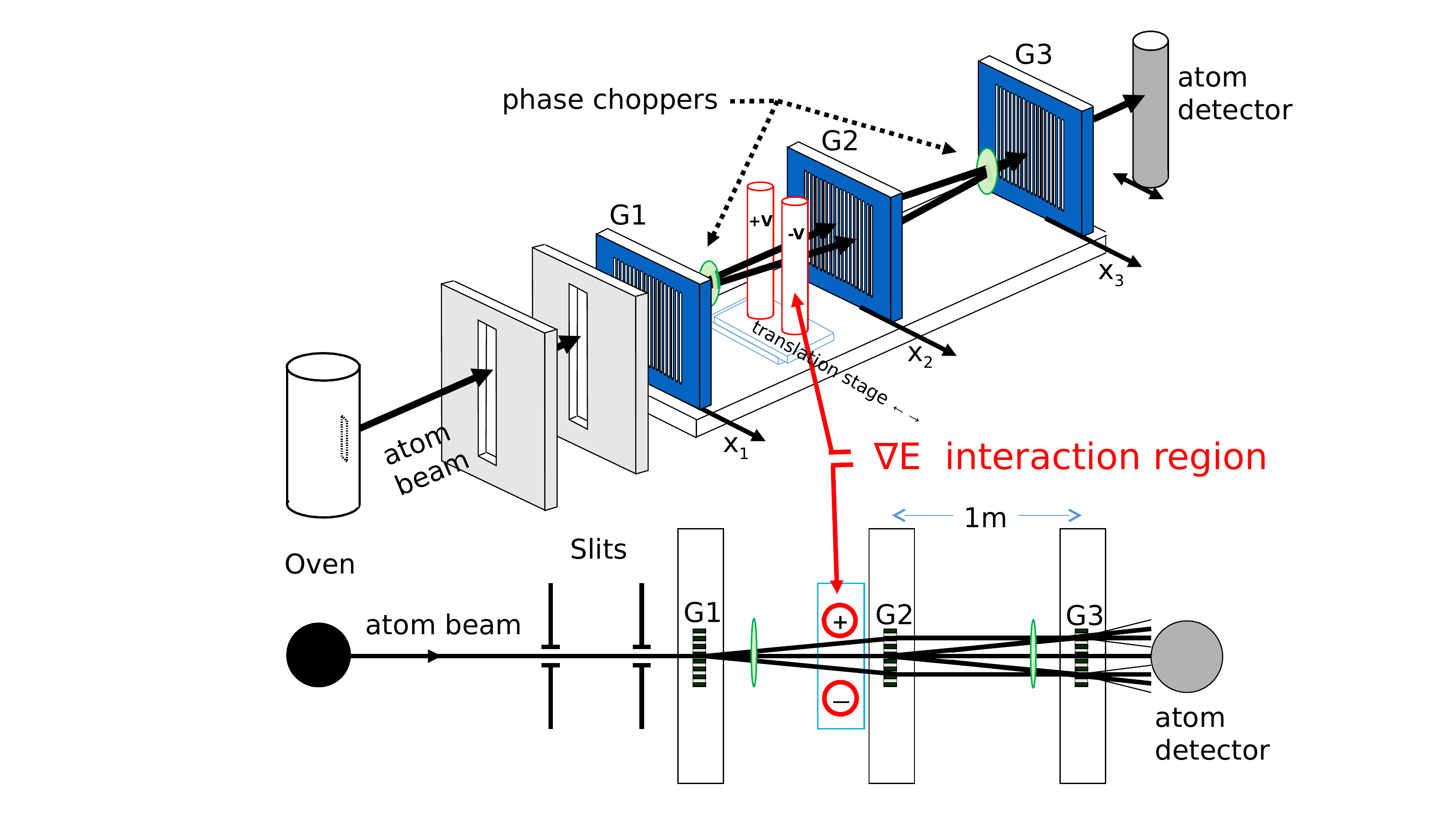}
\caption{\label{fig:apparatus} Diagram of the atom interferometer we used to measure the static polarizabilities of K, Rb, and Cs \cite{Gregoire2015}. A pair of cylindrical, oppositely-charged electrodes, indicated in red, induce phase shifts that depend on atoms' polarizabilities and the gradient of the produced electric field.
}
\end{figure}

\begin{figure}
\centering
\includegraphics[width=\linewidth,keepaspectratio]{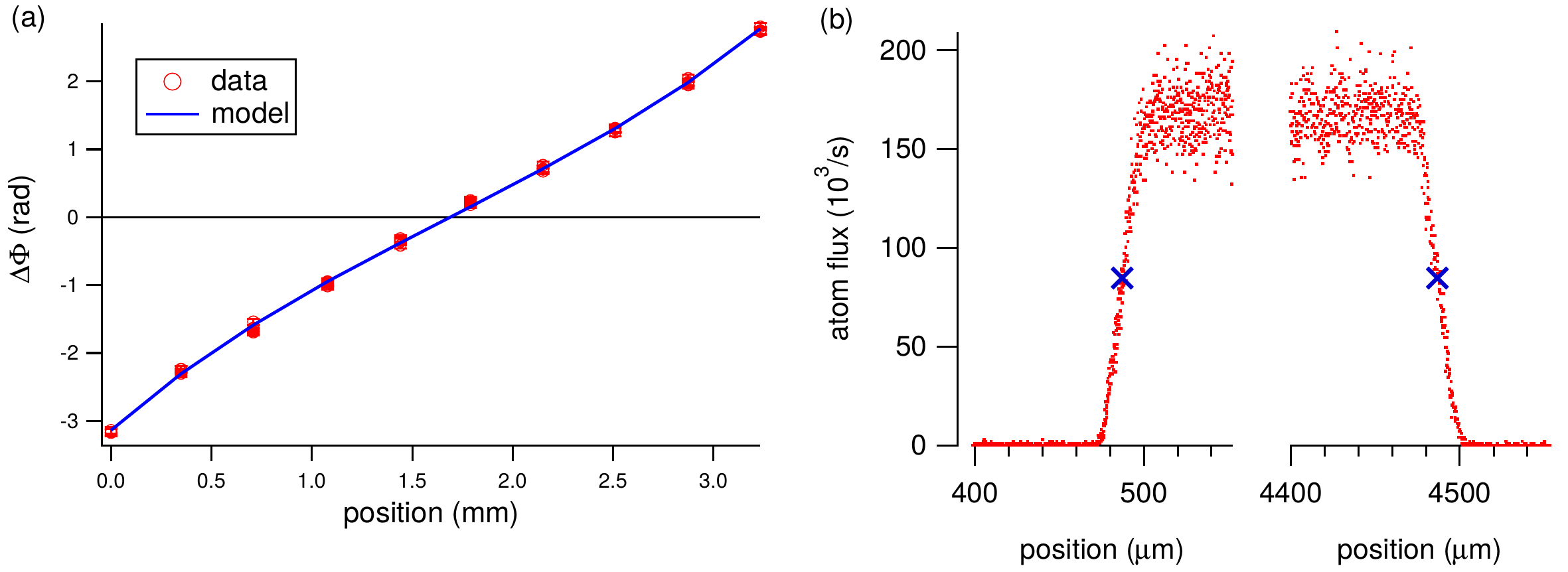}
\caption{\label{fig:apparatusData} (a): Induced phase shift in the interferometer vs the lateral position of the electrodes with respect to the beam. (b): Observed atom beam flux as a function of the electrodes' lateral position. We use these data to determine the distance between the electrodes.}
\end{figure}

We reduced the uncertainties in our most recent K, Rb, and Cs static polarizability measurements \cite{Gregoire2015} to 0.11\% by reducing the total systematic uncertainty from 0.15\% to 0.10\%.
In our experiment, we used cylindrical electrodes, indicated in red in \figref{fig:apparatus}, to induce phase shifts in our atom interferometer that are proportional to $\alpha(0)V^2$, the atoms' static polarizabilities times the square of the voltage difference between the electrodes. \figref{fig:apparatusData}a shows an example of how the induced phase shift changes as we move the electrodes laterally with respect to the beamline. 

We reduced the systematic uncertainty in our measurements from 0.15\% to 0.12\% by calibrating the voltage supplies connected to the electrodes to 36 ppm using a Vitrek 4700 high-accuracy voltmeter. Each electrode is held at its respective positive or negative voltage with respect to ground by its own power supply. We concluded that when we instructed the power supplies to output $\pm 6$ kV, both power supplies were actually supplying $\pm 6.0026(2)$ kV. Our results agreed with less-accurate calibration measurements of $\pm 6.003(3)$ we made earlier using a Fluke 287 multimeter and a Fluke 80k-40 high-voltage probe. At normal operating temperatures, our calibration measurements were completely reproducible to within the resolution of the Vitrek 4700.

The strength of the electric field gradient, and therefore the magnitude of the induced phase shift, also depends on the distance between the electrodes.
In the past, we measured that distance to be 1999.9(5) $\mu$m by sweeping the electrodes across the beamline and measuring the lateral positions at which the electrodes eclipsed the beam (see an example of these data in \figref{fig:apparatusData}b).
We found that scatter in our measurements was explained by misalignment of the collimating slits and detector. After correcting for this source of error, we measured the distance between electrodes to be 1999.7(2) $\mu$m, which further reduced our total systematic uncertainty from 0.12\% to 0.10\%. Our measurements did not change as a function of maximum atom flux, electrodes translation motor speed, atom beam $y$ position or vertical collimation, atom beam velocity, or atomic species. 

By themselves, the new values we measured for the electrodes' voltages and the distance between the electrodes changed our reported polarizabilities by +140 ppm and -140 ppm, respectively. Therefore, the polarizability values that we report are the same as those in \cite{Gregoire2015} but with smaller uncertainties. It is also worth noting that either of the $\pm 140$ ppm changes, by itself, would still not have been statistically significant.
These reduced total uncertainties are shown alongside the previously-reported values in Table \ref{tab:a0measurements}.

\section{Analysis of atom interferometry polarizability measurements \label{sec:pol2other}}

Table \ref{tab:a0measurements} lists  polarizability measurements made with atom interferometry.  For Li, Na, K, Rb, and Cs, atom interferometry has provided the best available measurements.  
Polarizability measurements made using other methods are reviewed in \cite{Amini2003,Mitroy2010, Schwerdtfeger2006a, Schwerdtfeger, Gould2005, Haynes2014}.

\begin{table}
\centering
\caption{Measurements of static polarizabilities $\alpha(0)$ made using atom or molecule interferometry.  References \cite{Ekstrom1995,Miffre2006} used a septum electrode and references \cite{Berninger2007, Holmgren2010, Gregoire2015} used electric field gradients to shift the phase of matter wave interference fringes.  Results are presented both in \AA$^3$ and atomic units (au). Values we use for analysis in this paper are in bold.}
\label{tab:a0measurements}
\begin{tabular}{lllll}
\toprule
\textbf{Atom or} & \multicolumn{2}{c}{\textbf{Polarizability}} & \textbf{Reference} &  \textbf{Uncertainty}   \\
\textbf{molecule}    &  \multicolumn{1}{c}{(\AA$^3$)}  & \multicolumn{1}{c}{(au)}  &   &      \\
\midrule
Li   & \textbf{24.33(16)} & \textbf{164.2(11)}  &      \cite{Miffre2006}  &  0.66\%          \\
\arrayrulecolor[gray]{0.8} \midrule
Na   & \textbf{24.11(8)}   &\textbf{162.7(5)} &     \cite{Ekstrom1995}  &  0.35\% \\
Na   & 24.11(18) & 162.7(12)  &    \cite{Holmgren2010}  &  0.75\% \\
\midrule
K    & 43.06(21) & 290.6(14)  &    \cite{Holmgren2010}  &  0.49\% \\
K    & 42.93(7) & 289.7(5)  &    \cite{Gregoire2015}  &  0.16\% \\
K    & \textbf{42.93(5)} & \textbf{289.7(3)}  &    this work  &  0.11\% \\
\midrule
Rb   & 47.24(21) & 318.8(14)  &    \cite{Holmgren2010}  &  0.44\% \\
Rb   & 47.39(8)  & 319.8(5)  &    \cite{Gregoire2015}  &  0.17\% \\
Rb   & \textbf{47.39(5)}  & \textbf{319.8(3)}  &    this work  &  0.11\% \\
\midrule
Cs   & 59.39(9)  & 400.8(6)  &    \cite{Gregoire2015}  &  0.15\% \\
Cs   & \textbf{59.39(6)}  & \textbf{400.8(4)}  &    this work  &  0.11\% \\
\midrule
C$_{60}$ & 88.9(52)  & 600(35)  &  \cite{Berninger2007} & 5.9\% \\
\midrule
C$_{70}$ & 108.5(65) & 732(44)  &  \cite{Berninger2007} & 6.5\% \\
\arrayrulecolor{black} \bottomrule
\end{tabular}                  
\end{table}

The original references \cite{Ekstrom1995, Holmgren2010, Gregoire2015} show how the polarizability measurements in Table \ref{tab:a0measurements} compare to theoretical predictions  \cite{Mitroy2010, Reinsch1976, Tang1976,Maeder1979,  Christiansen1982, Fuentealba1982, Muller1984, kundu1986,  Kello1993, Fuentealba1993, Marinescu1994, Dolg1996, Patil1997, Lim1999,  Safronova1999, Derevianko1999, Magnier2002, Mitroy2003, Safronova2004, Lim2005, Arora2007, Iskrenova-Tchoukova2007, Safronova2008, Safronova2011}.  In this article, we devote our attention to interpreting the atomic polarizability measurements in Table \ref{tab:a0measurements} in a systematic and tutorial manner.  In the rest of Section \ref{sec:pol2other} we show how to use these polarizability measurements to predit other atomic properties such as  oscillator strengths, lifetimes, matrix elements,  line strengths, and van der Waals $C_6$ coefficients, following procedures described earlier by Derevianko and Porsev \cite{Derevianko2002}, Amini and Gould \cite{Amini2003}, and Mitroy, Safronova, and Clark \cite{Mitroy2010} among others.  Then, in Section \ref{sec:testing_ar}, we use the polarizabilities in Table \ref{tab:a0measurements} to provide experimental constraints on the residual polarizabilities, $\alpha_r$, for each of the alklai atoms.


\subsection{Reporting oscillator strengths, lifetimes, matrix elements, and line strengths from static polarizabilities \label{sec:fADS}}

The dynamic polarizability, $\alpha(\omega)$,  of an atom in state $|i\rangle$ can be written as a sum over electric-dipole transition matrix elements $\langle k | e \vec{r} | i \rangle$, Einstein coefficients $A_{ik}$, oscillator strengths $f_{ik}$, or line strengths $S_{ik}$ as
\begin{align}
	\alpha(\omega) &= \frac{e^2}{m} \sum_{k\neq i} 
		\frac{f_{ik}}{\omega_{ik}^2 - \omega^2} 
	\label{eq:alphaf} \\
		\alpha(\omega) &= 2\pi\epsilon_0 c^3 \sum_{k\neq i} 
		\frac{A_{ki}\omega_{ik}^{-2}}{\omega_{ik}^2 - \omega^2} \frac{g_k}{g_i}
	\label{eq:alphaA} \\ 
	\alpha(\omega) &= \frac{2}{3 \hbar} \sum_{k\neq i} 
		\frac{ |\langle k | e \vec{r} | i \rangle|^2 \omega_{ik}}{\omega_{ik}^2 - \omega^2}
	\label{eq:alphamu} \\
	\alpha(\omega) &= \frac{1}{3 \hbar} \sum_{k\neq i} 
		\frac{S_{ik}\omega_{ik}}{\omega_{ik}^2 - \omega^2} 
	\label{eq:alphaS}
\end{align}
where $e$ and $m$ are the charge and mass of an electron,  $\omega_{ik} = (E_k - E_i)/\hbar$ are resonant frequencies for excitation from state $|i\rangle$ to state $|k\rangle$,  and $g_k = 2J_k+1$ is the degeneracy of state $|k\rangle$. The squares of electric dipole transition matrix elements  $|\langle k|e\vec{r} |i\rangle |^2$, or equivalently $3|\langle k|e\vec{x}|i \rangle|^2$, are related to the reduced dipole matrix elements (denoted with double bars) by  $|\langle k  \| e\vec{r} \| i \rangle|^2 = |D_{ik}|^2 = \sum_{m_k,m_i}|\langle k | e\vec{r} | i \rangle|^2 = |\langle k|e\vec{r}|i\rangle|^2g_i$ using the Wigner-Eckart theorem. For ground state alkali atoms, line strength $S_{ik} = |D_{ik}|^2$.  

\color{black}

The expressions for polarizability $\alpha(0)$ in Eqs.~(\ref{eq:alphaf}) - (\ref{eq:alphaS}) each have dimensions of $4\pi\epsilon_0$ times volume, as expected from the definitions $\vec{p}=\alpha \vec{E}$ and $U=-\frac{1}{2}\alpha |\vec{E}|^2$, where $\vec{p}$ is the induced dipole moment and $U$ is the energy shift (Stark shift) of an atom in an electric field $\vec{E}$. When polarizability is reported in units of volume (typically \AA$^3$ or $10^{-24}$ cm$^3$) it is implied that one can multiply by $4\pi\epsilon_0$ to get polarizability in SI units. The atomic unit (au) of polarizability, $e^2 a_0^2/E_h$, is equivalent to $a_0^3 \times 4\pi\epsilon_0$, where $a_0$ is the Bohr radius, and $E_h$ is a Hartree. Since $(4\pi\epsilon_0) = 1$ in au, polarizability is naturally expressed in atomic units of volume of $a_0^3$  (and for reference $a_0^3 = 0.148185$ \AA$^3$).

Since the principal D1 and D2 transitions of alkali metal atoms (denoting the $ns$-$np_{1/2}$ and $ns$-$np_{3/2}$  transitions respectively, where $n$=6 for Cs, $n$=5 for Rb,  $n$=4 for K, $n$=3 for Na, and  $n$=2 for Li), account for over 95\% of those atoms' static polarizabilities \cite{Derevianko1999}, it is customary to decompose polarizability as
\begin{align}
\alpha(\omega) =  \alpha_{p}(\omega) + \alpha_r(\omega)
\label{eq:alpha_p_r}
\end{align}
where $\alpha_{p}(\omega)$ represents the contribution from the principal transitions and $\alpha_r(\omega)$ is the residual polarizability due to all other excitations.  
The residual polarizability itself can be further decomposed as
\begin{align}
\alpha_r(\omega) = \alpha_{v'}(\omega) + \alpha_{\mathrm{core}}(\omega) + \alpha_{cv}(\omega)
\label{eq:alpha_r_sum}
\end{align}
where $\alpha_{v'}$ is due to higher $ns_{1/2} - n'p_J$ excitations of the valence electron with $n'>n$, $\alpha_{\mathrm{core}}$ is the polariability due to the core electrons, and $\alpha_{cv}$ is due to correlations between core and valence electrons.   Sometimes the notation  $\alpha_{\mathrm{tail}}$ is used to denote a  subset of $\alpha_{v'}$ with $n'>(n+3)$  \cite{Safronova2006}, or  $n'>(n+5)$  \cite{Safronova1999}, or an even higher cutoff such as $n'>26$  \cite{Safronova2008}. 
 
Using the decomposition in \eqnref{eq:alpha_p_r} we can rewrite Eqs.~(\ref{eq:alphaf})-(\ref{eq:alphaS}) for static ($\omega=0$) polarizabilities:
\begin{align}
	\alpha(0) &= \frac{e^2}{m} 
	\left[  \frac{ f_{D1}}{\omega_{D1}^2} + \frac{f_{D2}}{\omega_{D2}^2} \right] + \alpha_{r}(0) 
	\label{eq:a0f} \\
	\alpha(0) &= 2\pi\epsilon_0 c^3 
	\left[  \frac{\tau_{1/2}^{-1}}{\omega_{D1}^4} + 2 \frac{\tau_{3/2}^{-1}}{\omega_{D2}^4} \right] + \alpha_{r}(0) 		\label{eq:a0A} \\
	\alpha(0) &= \frac{1}{3\hbar } 
	\left[   \frac{|D_{D1}|^2}{\omega_{D1}} +  \frac{|D_{D2}|^2}{\omega_{D2}} \right] + \alpha_{r}(0) 
	\label{eq:a0D} \\
	\alpha(0) &= \frac{1}{3\hbar} 
	\left[  \frac{ S_{D1}}{\omega_{D1}} + \frac{S_{D2}}{\omega_{D2}} \right] + \alpha_{r}(0) 
	\label{eq:a0S} 
\end{align}
\eqnref{eq:a0A} is written in terms of lifetimes $\tau_k^{-1} = \sum_{i} A_{ki}$, rather than Einstein $A$ coefficients because alkali metal atom $np_J$ states decay with a branching ratio of 100\% to their respective ground $ns_{1/2}$ states. To support our analysis of polarizabilites here in Section \ref{sec:pol2other} we use theoretically caclulated values of residual static polarizabilities $\alpha_r(0)$ = 2.04(69) au for Li, $\alpha_r(0)$ = 1.86(12) au for Na, $\alpha_r(0)$ = 6.26(33) au for K, $\alpha_r(0)$ = 10.54(60) au for Rb, all from Savronova \etal \cite{Safronova2006}, and $\alpha_r(0)$ = 16.74(11) au for Cs from Derevianko \etal \cite{Derevianko2002}.   Table \ref{tab:ar} in Appendix \ref{sec:ar} lists these and several other published values for $\alpha_{\mathrm{core}}(0)$, $\alpha_{v'}(0)$, $\alpha_{cv}(0)$ and $\alpha_r(0)$.  

Since $\omega_{D1}$ and $\omega_{D2}$ are well known \cite{NIST}, we can further use Eqs.~\eqref{eq:a0f}-\eqref{eq:a0S} to derive expressions for $|D_{ik}|^2$, $\tau_k$, and $f_{ik}$ in terms of $\alpha(0)$, $\alpha_r(0)$, and a ratio of line strengths $R$:
\begin{align}
	f_{D1} &= \frac{\left[\alpha(0)-\alpha_r(0)\right]}{\left(\frac{e^2}{m\omega_{D1}^2}\right)}\left(\frac{1}{1+R\frac{\omega_{D1}}{\omega_{D2}}}\right)
	\label{eq:f10a} \\
	f_{D2} &= \frac{\left[\alpha(0)-\alpha_r(0)\right]}{\left(\frac{e^2}{m\omega_{D1}^2}\right)}\left(\frac{R}{\frac{\omega_{D2}}{\omega_{D1}}+R}\right)
	\label{eq:f20a} \\
	\tau_{1/2} &= \frac{2\pi\epsilon_0c^3\omega_{D1}^{-3}}{\left[\alpha(0)-\alpha_r(0)\right]}\left(\frac{1}{\omega_{D1}}+\frac{R}{\omega_{D2}}\right)
	\label{eq:A10a} \\
	\tau_{3/2} &= \frac{2\pi\epsilon_0c^3\omega_{D2}^{-3}}{\left[\alpha(0)-\alpha_r(0)\right]}\left(\frac{2}{R\omega_{D1}}+\frac{2}{\omega_{D2}}\right)
	\label{eq:A20a} \\
	|D_{D1}|^2 &= S_{D1} = \left[\alpha(0)-\alpha_r(0)\right]\left(\frac{3\hbar}{\frac{1}{\omega_{D1}}+\frac{R}{\omega_{D2}}}\right)
	\label{eq:d10a} \\
	|D_{D2}|^2 &= S_{D2} = \left[\alpha(0)-\alpha_r(0)\right]\left(\frac{3\hbar}{\frac{1}{R\omega_{D1}}+\frac{1}{\omega_{D2}}}\right)
	\label{eq:d20a}
\end{align}
where $R$ is defined as
\begin{align}
	R \equiv 
	\frac{S_{D2}}{S_{D1}} =  
	\frac{|D_{D2}|^2}{|D_{D1}|^2} = 
	\frac{ f_{D2} }{ f_{D1} }  
	\frac{\omega_{D1} }{ \omega_{D2}} =  
	2 \frac{\tau_{1/2} }{ \tau_{3/2}} \left(\frac{\omega_{D1} }{ \omega_{D2}} \right)^3 
	\label{eq:Rdef}
\end{align}

To support our analysis of polarizabilities, we will use $R$= 2.0000 for Li  inferred from \cite{Johnson2008},  $R$=1.9994(37) for Na \cite{Volz1996}, $R$=1.9976(13) for K \cite{Trubko2016}, $R$=1.99219(3) for Rb \cite{Leonard2015} , and $R$= 1.9809(9) for Cs \cite{Rafac1998}.   It is noteworthy that references \cite{Holmgren2012, Leonard2015, Trubko2016} determined $R$ experimentally using atom interferometry measurements of tune-out wavelengths. 



\begin{table}
\centering
\caption{Atomic properties inferred from atom interferometry measurements \cite{Ekstrom1995,Miffre2006,Gregoire2015}.  Reduced matrix elements $D_{D1} = \langle np_{1/2} \| r \| ns_{1/2} \rangle $ and $D_{D2} = \langle np_{3/2} \| r \| ns_{1/2} \rangle $, lifetimes $\tau_{np1/2}$ and $\tau_{np3/2}$, oscillator strengths $f$, and line strengths $S$  shown here are inferred from measurements of  polarizabilities, $\alpha(0)$ shown in Table \ref{tab:a0measurements} using Eqns (\ref{eq:f10a}) - (\ref{eq:d20a}).  Subscripts D1 and D2 refer to the $ns$-$np_{1/2}$ and $ns$-$np_{3/2}$ transitions respectively, where $n$=6 for Cs, $n$=5 for Rb, $n$=4 for K, $n$=3 for Na and $n$=2 for Li.  Uncertainty budget components $\delta_{\alpha}$, $\delta_R$, and  $\delta_{\alpha_r}$ come from the uncertainties in $\alpha(0)$ \cite{Ekstrom1995,Miffre2006,Gregoire2015} (see Table \ref{tab:a0measurements}), $R$ \cite{Johnson2008, Volz1996, Trubko2016, Leonard2015, Rafac1998}, and $\alpha_r(0)$ \cite{Safronova2006, Derevianko2002} (see Table \ref{tab:ar}).  The resulting uncertainties for $D_{D1}$, $D_{D2}$, $\tau_{np1/2}$, $\tau_{np3/2}$, $f_{D1}$, $f_{D2}$, $S_{D1}$, and $S_{D2}$  are reported presuming possible errors $\delta_{\alpha}$, $\delta_R$, and  $\delta_{\alpha_r}$  are uncorrelated. The symbol (-) indicates an uncertainty $<1$ in the least significant digit.}
\label{tab:ff}
\begin{tabular}{lllllllllll}
\toprule
atom        & $D_{D1}$  (au) &  $\delta_{\alpha}$   &  $\delta_R$  &  $\delta_{\alpha_r}$   &  \,\,\,\,\,\,  
            & $D_{D2}$ (au) &  $\delta_{\alpha}$   &  $\delta_R$  &  $\delta_{\alpha_r}$   \\
\midrule
   Li &  3.318(13) & (11) & (-) &  (7)   & &  4.693(19) &  (16) & (-) & (10)     \\
  Na &  3.527(6)  & (6)  & (2) &   (1)  & &  4.987(8)  &  (8)    &  (2) &  (2)   \\
  K  &  4.101(4)   & (3)  & (1) &   (2)   & &  5.800(5)  &  (4)   &  (1) &  (3)  \\
  Rb &  4.239(5)  & (3)  & (-) &   (4)   & &  5.989(7)  &  (4)   &  (-) &  (6)  \\
  Cs &  4.508(3)  & (3)  & (1) &   (1)   & &  6.345(4)  &  (4)   &  (-) &  (1)  \\
\toprule
atom  & $\tau_{1/2}$  (ns) &  $\delta_{\alpha}$   &  $\delta_R$     &  $\delta_{\alpha_r}$  &  
      & $\tau_{3/2}$ (ns) &  $\delta_{\alpha}$   &  $\delta_R$     &  $\delta_{\alpha_r}$     \\
\midrule 
  Li &   27.08(21) & (18)  &  (-)  &  (11) & &   27.08(21) &  (18) & (-)  & (11) \\
  Na &   16.28(6) & (5)   & (2)  &  (3)   & &   16.24(5) & (5)   & (1)  &  (1)      \\
  K  &   26.80(4)  & (3)   & (1)  &  (3)    & &   26.45(4)  & (3)   & (1)  &  (3)    \\
  Rb &   27.60(6) & (3)   & (-)  &  (5)   & &   26.14(6)  & (3)   & (-)  &  (5)     \\
  Cs &   34.77(5) & (5)   & (1)  &  (1)   & &   30.37(5)  & (4)   & (1)  &  (-)     \\
\toprule
atom & $f_{D1}$  &  $\delta_{\alpha}$  & $\delta_R$  &  $\delta_{\alpha_r}$   & \,\,\,\,\,\,  
     & $f_{D2}$  &  $\delta_{\alpha}$  & $\delta_R$  &  $\delta_{\alpha_r}$  \\
\midrule
  Li & 0.2492(20)   & (17)    &  (-)  & (11)    &&     0.4985(39)  & (33)  &  (-)   &  (39)        \\
  Na & 0.3203(12)  & (11)  & (4)  & (2)   & &  0.6410(23)  & (22)  & (4)  & (5)    \\
  K  & 0.3317(6)  & (4)   & (1)  & (4)      & &  0.6665(11)  & (8)  & (1)  & (7)   \\
  Rb & 0.3438(8)  & (4)   & (-)  & (7)    & & 0.6982(17)  & (9)  & (-) & (14)  \\
  Cs & 0.3450(5)  & (5)   & (1)  & (1)     & &  0.7174(10)  & (10)  & (1)  & (2)  \\
\toprule
atom     & $S_{D1}$ (au) &  $\delta_{\alpha}$   &  $\delta_R$     &  $\delta_{\alpha_r}$   & \,\,\,\,\,\,  
         & $S_{D2}$ (au) &  $\delta_{\alpha}$   &  $\delta_R$     &  $\delta_{\alpha_r}$     \\
\midrule
  Li &   11.01(9)    &  (8)    &    (-)     & (5)   &&   22.02(17)      &  (15)      &  (-)      & (9)         \\
  Na &  12.44(5)  & (4)   & (2)   &  (1)  & & 24.87(8)  &  (8)   &  (2)   &  (2)  \\
  K  &  16.82(3)  & (2)   & (1)   &  (2)   & & 33.64(6)  &  (4)   &  (1)   &  (4)   \\
  Rb &  17.97(4)  & (3)   & (-)   &  (3)  & & 35.87(8)  &  (4)   &  (-)   &  (7)    \\
  Cs &  20.32(3)  & (3)   & (1)   &  (1)  & & 40.26(5)  &  (5)   &  (1)   &  (1)  \\
\bottomrule
\end{tabular}
\end{table}

Table \ref{tab:ff} shows principal transition matrix elements, lifetimes, line strengths, and oscillator strengths 
inferred from polarizability measurements using Eqns.~\eqref{eq:f10a}-\eqref{eq:Rdef}. 
Our inferred lifetimes for K, Rb, and Cs are based on $\alpha(0)$ measurements with 0.11\% uncertainty, yet our derived lifetimes have slightly larger uncertainty. In the case of Li, Na, K and Rb, this is because roughly half of the total uncertainty comes from uncertainty in $\alpha_r(0)$, whereas for Cs the uncertainties in $\tau$ are dominated by contributions from uncertainty in $\alpha(0)$. 
Because there have been many high precision measurements of alkali metal principal transition lifetimes, it is useful to compare our derived lifetimes to those measurements. 
Our derived K and Rb lifetimes agree well with and have comparable uncertainty to those measured by Volz \etal \cite{Volz1996}, Wang \etal \cite{Wang1997,Wang1997a}, and Simsarian \etal \cite{Simsarian1998}. 
Because the $\alpha(0)$ measurements used to derive the Li and Na lifetimes in Table \ref{tab:ff} are less precise, our inferred Na lifetimes have about twice the uncertainty (about 0.4\%) of measurements by Volz \etal \cite{Volz1996}, and our inferred Li lifetimes have much greater uncertainty than measurements by Volz \etal \cite{Volz1996} and McAlexander \etal \cite{McAlexander1996}.

\begin{figure}
\centering
\includegraphics[width=0.8\linewidth,keepaspectratio]{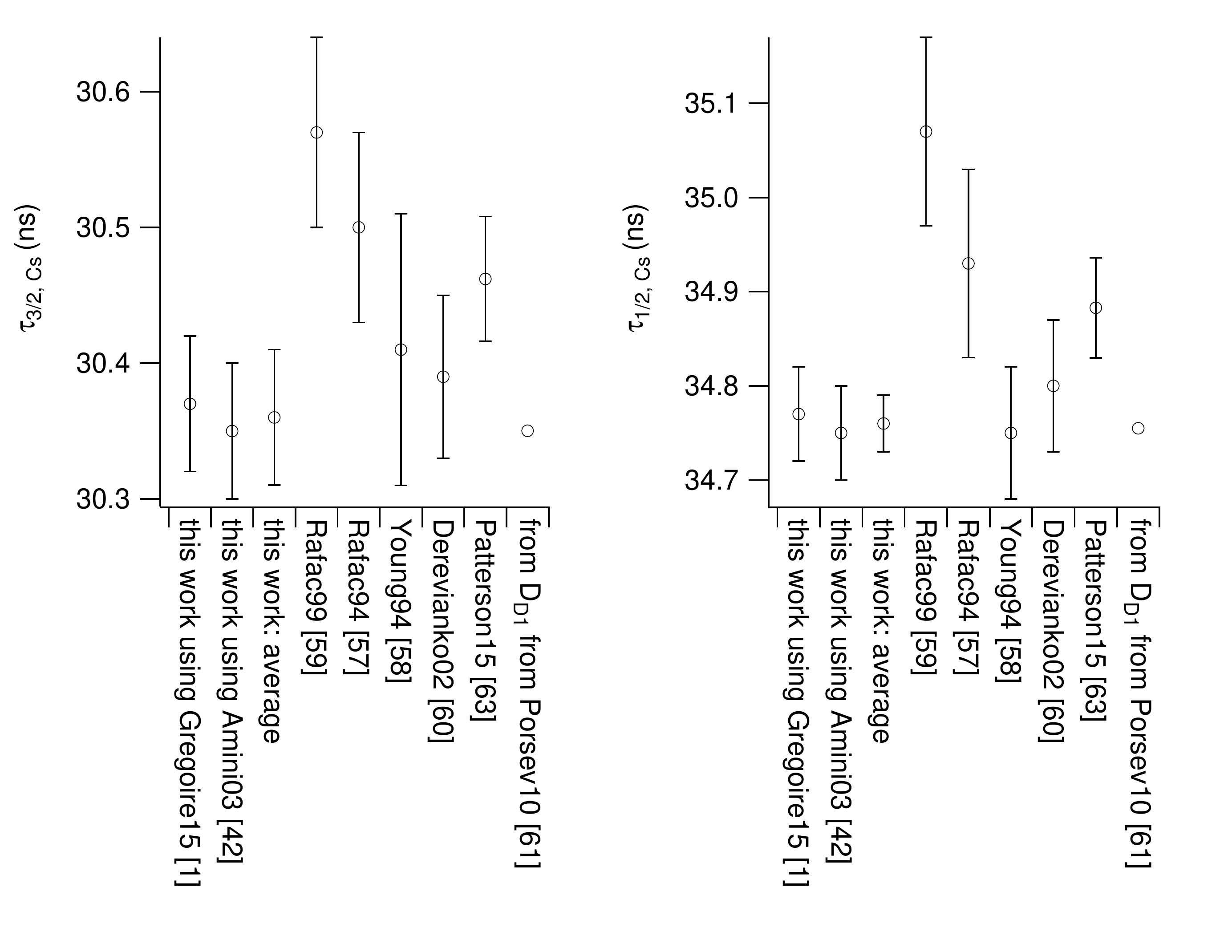}
\caption{\label{fig:tau_cs} Comparisons of Cs principal transition lifetimes inferred from polarizability measurements \cite{Amini2003,Gregoire2015} and theoretical $\arcs$ \cite{Derevianko2002}, direct measurements \cite{Rafac1999,Rafac1994,Young1994,Derevianko2002b}, and a combination of Porsev \etal's calculated $|D_{D1}|$ \cite{Porsev2010} and $R$ \cite{Rafac1998}.
}
\end{figure}

For Cs, the lifetimes we report in Table \ref{tab:ff} for this work have an uncertainty of less than 0.15\%, which is slightly smaller than the uncertainty of four previous high-precision determinations of the Cs $6p_J$ state lifetimes \cite{Rafac1994, Young1994, Rafac1999, Derevianko2002b}.
Table \ref{tab:tau_cs} and \figref{fig:tau_cs} show how our lifetime results are consistent with \cite{Young1994,Derevianko2002b} but  differ from lifetimes reported in \cite{Rafac1994, Rafac1999}. Our results deviate by 1.5$\sigma$ from $\tau_{1/2}$ found in \cite{Rafac1994} and by 3$\sigma$ from $\tau_{1/2}$ in \cite{Rafac1999}, where $\sigma$ for the deviations here refers to the combined uncertainty (added in quadrature) for the experiments.   Comparing the sum of line strengths ($S_{D1} + S_{D2}$), a quantity that is mostly independent of $R$, provides a similar conclusion: our results are consistent with \cite{Young1994} and \cite{Derevianko2002b} but differ by two and three $\sigma$ from \cite{Rafac1994} and \cite{Rafac1999}.

Because the two recent measurements of $\acs$ by Gregoire \etal \cite{Gregoire2015} and Amini and Gould \cite{Amini2003} were made using very different methods, we combine these measurements using a weighted average in order to report a value for $\tau_{6p_{1/2},\mathrm{Cs}}$ with even smaller (0.03 ns) uncertainty in Table \ref{tab:tau_cs}. We note that, due to the uncertainty in $R$ and $\alpha_r(0)$, the uncertainty in $\tau_{6p_{1/2},\mathrm{Cs}}$ would still be 0.01 ns even if the polarizability measurements had no uncertainty.

The Cs $|D_{D1}|$ value calculated \emph{ab initio} by \cite{Porsev2010} is also consistent with our results for $|D_{D1}|$.   Since our results come from independent measurements of $\alpha(0)$ and $R$, combined with theoretical values for $\alpha_r(0)$ , the agreement between our result for $|D_{D1}|$ with the that of Derevianko and Porsev \cite{Derevianko2002b,Porsev2010} adds confidence to their analysis of atomic parity violation \cite{Porsev2009,Porsev2010}.

\begin{table}
\centering
\caption{Cesium $6p_J$ lifetimes ($\tau_J$) from several references, tabulated here for comparison.   The $\tau_{2/1}$ and $\tau_{3/2}$ values that we report using $\alpha(0)$ measured by atom interferometry \etal \cite{Gregoire2015} (combined with values of $\alpha_r(0)$ \cite{Porsev2010} and $R$ \cite{Rafac1998}) are reproduced from Table \ref{tab:ff}.   Similar comparisons appear in Table II of \cite{Amini2003} and Table I of \cite{Patterson2015}. }
\label{tab:tau_cs}
\begin{tabular}{lll}
\toprule
$\tau_{1/2}$  (ns) &  $\tau_{3/2}$ (ns)  &  Method and Reference(s)  \\
\midrule
34.77(5) & 30.37(5) & this work using $\alpha(0)$ from atom interferometry \cite{Gregoire2015} \\
34.75(5) & 30.35(5)   &  this approach using $\alpha(0)$ from \cite{Amini2003}  \\
34.76(3) & 30.36(3)   &  this approach using $\alpha(0)$ from both \cite{Gregoire2015} and \cite{Amini2003}  \\
\midrule
35.07(10)  &  30.57(7)      &  \cite{Rafac1999} Rafac 1999  \\
34.93(10)  &  30.50(7)      &  \cite{Rafac1994} Rafac 1994  \\
34.75(7)    &  30.41(10)    &  \cite{Young1994} Young 1994  \\
34.80(7)   &  30.39(6)       &  \cite{Derevianko2002b} Derevianko 2002 \\
34.883(53)&  30.462(46)   & $\tau_{3/2}$ from \cite{Patterson2015}, combined with $R$  \cite{Rafac1998} to infer $\tau_{1/2}$ \\
34.755       &   30.3502     &  from $D_{D1}$ calculation by \cite{Porsev2010}, combined with $R$ \cite{Rafac1998} to infer $\tau_{3/2}$ \\
\bottomrule
\end{tabular}
\end{table}

\subsection{Deriving van der Waals coefficients from polarizabilities \label{sec:derivingC6}}

Since polarizability determines the strengths of van der Waals (vdW) potentials, we can also use measurements of $\alpha(0)$ to improve predictions for atom-atom interactions. Two ground-state atoms have a van der Waals interaction potential
\begin{align}
    U = -\frac{C_6}{r^6} - \frac{C_8}{r^8} - \frac{C_{10}}{r^{10}} + ...
    \label{eq:Uvdw}
\end{align}
where $r$ is the inter-nuclear distance and $C_6$, $C_8$, and $C_{10}$ are dispersion coefficients that can be predicted  based on $\alpha(0)$ measurements.  For long-range interactions in the absence of retardation (i.e. for $a_0 \ll r \ll c/\omega_{D2}$), the $C_6$ term is most important.   The $C_6$ coefficient for homo-nuclear atom-atom vdW interactions depends on dynamic polarizability as
\begin{align}
	C_6 = \frac{3\hbar}{\pi} \int_0^\infty \left[ \alpha(i \omega) \right]^2 d\omega
	\label{eq:c6int}
\end{align}
Even though $\hbar = 1$ in au, we write $\hbar$ explicitly in Eqn.~(\ref{eq:c6int}) to emphasize that the dimensions of $C_6$ are energy $\times$ length$^6$.

The London result of $C_6 = (3/4) \hbar \omega_0 \alpha(0)^2$ can be found from Eqn.~(\ref{eq:c6int}) by using Eqn.~(\ref{eq:alphaf}) for $\alpha(i\omega)$ with a single term in the sum to represent an atom as a single oscillator of frequency $\omega_0$ with static polarizability $\alpha(0)$.  However, calculating $C_6$ gets more difficult for atoms with multiple oscillator strengths. In light of this complexity, we instead use the decomposition in \eqnref{eq:alpha_p_r} to express $C_6$ as
\begin{align}
	C_6 &= \frac{3\hbar}{\pi} \int_0^\infty \left[ \alpha_{p}(i \omega) + \alpha_{r}(i \omega) \right]^2 d\omega
	\nonumber \\
	&= \frac{3\hbar}{\pi} \int_0^\infty \left[ \alpha_p(i \omega) \right]^2 d\omega
	+ \frac{6\hbar}{\pi} \int_0^\infty \alpha_p(i \omega) \alpha_r(i \omega) d\omega
	+ \frac{3\hbar}{\pi} \int_0^\infty \left[ \alpha_r(i \omega) \right]^2 d\omega
	\label{eq:c6int_exp_general}
\end{align}
Because of the cross term, the integration over frequency, and the way $\alpha(i\omega)$ remains relatively constant until ultraviolet frequencies,              $\alpha_r$ is significantly more important for $C_6$ than for $\alpha(0)$. Contributions from $\alpha_r$ account for 15\% of $C_{6}$ whereas $\alpha_r$ contributes only 4\% to $\alpha(0)$ for Cs, as pointed out by Derevianko \etal \cite{Derevianko1999}.  

\begin{figure}
\centering
\includegraphics[width=\linewidth*2/3,keepaspectratio]{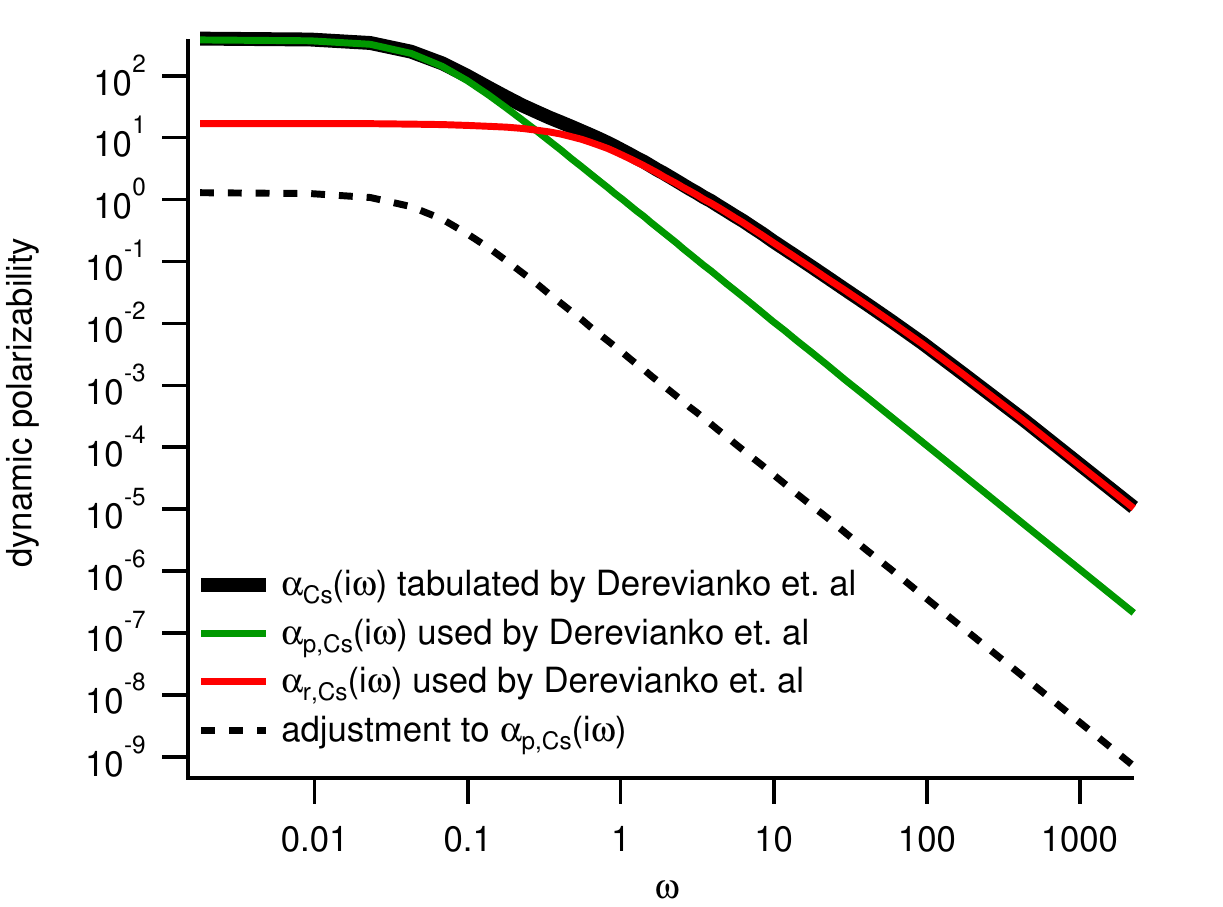}
\caption{\label{fig:alpha_cs} Cesium atom $\alpha_{tab}(i\omega)$ tabulated in \cite{Derevianko2010} (black line), and its decomposition into calculated $\alpha_p(i\omega)$ based on $\tau_{1/2}$ and $\tau_{3/2}$ values \cite{Volz1996} with Eq.~(\ref{eq:alphaA}) (green line) and the residual $\alpha_r(i\omega)$ (red line). The black dotted line represents the adjustment to $\alpha_p(i\omega)$ when we substitute $\alpha(0)$ measurement into \eqnref{eq:c6int_pfactored}. }
\end{figure}

The fact that $C_6$ and $\alpha(0)$ depend on $\alpha_r$ in different ways [compare Eqns.~(\ref{eq:alpha_p_r}) and (\ref{eq:c6int_exp_general})] suggests that it is possible to determine $\alpha_r(0)$ based on independent measurements of $C_6$ and $\alpha(0)$.  We will explore this in Section \ref{sec:testing_ar}.  First, we want to demonstrate how to use experimental $\alpha(0)$ measurements and theoretical $\alpha_r(i\omega)$ spectra to improve predictions of $C_6$ coefficients. For this we begin by factoring $\alpha_p(0)$ out of the $\alpha_p(i\omega)$ term in the integrand of \eqnref{eq:c6int_exp_general} to get
\begin{align}
	C_6 = \frac{3\hbar}{\pi} \int_0^\infty 
	\left[ \alpha_p(0)\frac{\alpha_p(i \omega)}{\alpha_p(0)} + \alpha_{r}(i \omega) \right]^2 
	d\omega 
	\label{eq:c6int_pfactored}
\end{align}
where the spectral shape function
\begin{align}
	\frac{\alpha_p(i\omega)}{\alpha_p(0)} = 
	\frac{
		\frac{1}{\omega_{D1}^2+\omega^2} + \frac{R\frac{\omega_{D1}}{\omega_{D3}}}{\omega_{D3}^2+\omega^2}
	}{
		\frac{1}{\omega_{D1}^2} + \frac{R\frac{\omega_{D1}}{\omega_{D3}}}{\omega_{D3}^2}
	}
	\label{eq:pshape}
\end{align}
uses $R$ defined in \eqnref{eq:Rdef}. We are now able to calculate $C_6$ using our choice of $\alpha_p(0)$, which we can relate to static polarizability measurements via $\alpha_p(0) = \alpha(0) - \alpha_r(0)$.  The formula for $C_6$ can then be written as
\begin{align}
	C_6 = \frac{3\hbar}{\pi} \int_0^\infty 
	\left[ [\alpha(0)-\alpha_r(0)]\frac{\alpha_p(i \omega)}{\alpha_p(0)} + \alpha_{r}(i \omega) \right]^2 
	d\omega
	\label{eq:c6int_pfactored}
\end{align}
To use \eqnref{eq:c6int_pfactored} to infer values of $C_6$ from our static polarizability measurements, one still needs to know $\alpha_r(i\omega)$ and $\alpha_r(0)$. Derevianko \etal calculated and tabulated values $\alpha_{tab}(i\omega)$ in \cite{Derevianko2010} of polarizability for all the alkali atoms, where the principal component $\alpha_p(i\omega)$ was calculated using experimental lifetime measurements by Volz and Schmoranzer \cite{Volz1996}  for Li, Na, K, and Rb and by Rafac \etal \cite{Rafac1994}  for Cs.  Therefore, we know that the residual component $\alpha_r(i\omega)$ of Derevianko \etalnospace's tabulated values of $\alpha_{tab}(i\omega)$ is
\begin{align}
	\alpha_r(i\omega) = \alpha_{tab}(i\omega) - 
	2 \pi \epsilon_0 c^3 \left[
		\frac{\tau_{1/2}^{-1}\omega_{D1}^{-2}}{\omega_{D1}^2-\omega^2} +
		2 \frac{\tau_{3/2}^{-1}\omega_{D2}^{-2}}{\omega_{D2}^2-\omega^2}
	\right]
	\label{eq:ar_tab}
\end{align}
\figref{fig:alpha_cs} shows an example of how  $\alpha_{tab}(i\omega)$ for Cs tabulated by Derevianko \etal \cite{Derevianko2010} can be decomposed into principal and residual parts.  \figref{fig:alpha_cs} also shows the small adjustment to $\alpha_p(i\omega)$ that can be recommended based on measurements of $\alpha(0)$.  In essence, this procedure makes the assumption that any deviation between the measured and the tabulated \cite{Derevianko2010} values of static polarizability are due to an error in the $\alpha_p$ part of the tabulated values, and that the $\alpha_r(i\omega)$ component of the tabulated values is correct.  To assess the impact of this assumption, next we examine how uncertainty in $\alpha_r(i\omega)$ propagates to uncertainty in $C_6$.

Equation (\ref{eq:c6int_pfactored}) shows how $C_6$ calculations depend on $\alpha_r(0)$ and $\alpha(i\omega)$ with opposite signs.  This helps explain why uncertainty in $\alpha_r$ propagates to uncertainty in $C_6$ with a somewhat reduced impact.   For example, if $\alpha_r$ accounts for 15\% of $C_6$, and $\alpha_r$ itself has an uncertainty of 5\%, one might naievely expect that uncertainty in $C_6$ due to uncertainty in $\alpha_r$ would be 0.75\%.  However, using Eqn.~(\ref{eq:c6int_pfactored}) one can show that the uncertainty in $C_6$ is smaller  (only 0.48\% due to $\alpha_r$).  To explain this, if a  theoretical  $\alpha_r$ is incorrect, say a bit too high, then when we subtract this from the measured $\alpha(0)$ we will deduce an $\alpha_p(0)$ that is too small, and the error from this contribution to $C_6$ has the opposite sign from the error caused by adding back $\alpha_r(i\omega)$ in  \eqref{eq:c6int_pfactored}.

We can also rewrite \eqnref{eq:c6int_pfactored} by adding and subtracting the tabulated $\alpha_p(i\omega)$ so that $C_6$ depends explicitly only on the measured and  tabulated (total) polarizabilities.
\begin{align}
	C_6 = \frac{3\hbar}{\pi} \int_0^\infty
	\left[ [\alpha(0)-\alpha_{tab}(0)]\frac{\alpha_p(i \omega)}{\alpha_p(0)}+\alpha_{tab}(i\omega) \right]^2
	d\omega
	\label{eq:c6int_nop}
\end{align}
where $\alpha_{tab}(i\omega)$ and $\alpha_{tab}(0)$ refer to values tabulated by Derevianko \etal   This way  $C_6$ does not explicitly depend on              $\alpha_r$.

\begin{figure}
\includegraphics[width=0.8\linewidth,keepaspectratio]{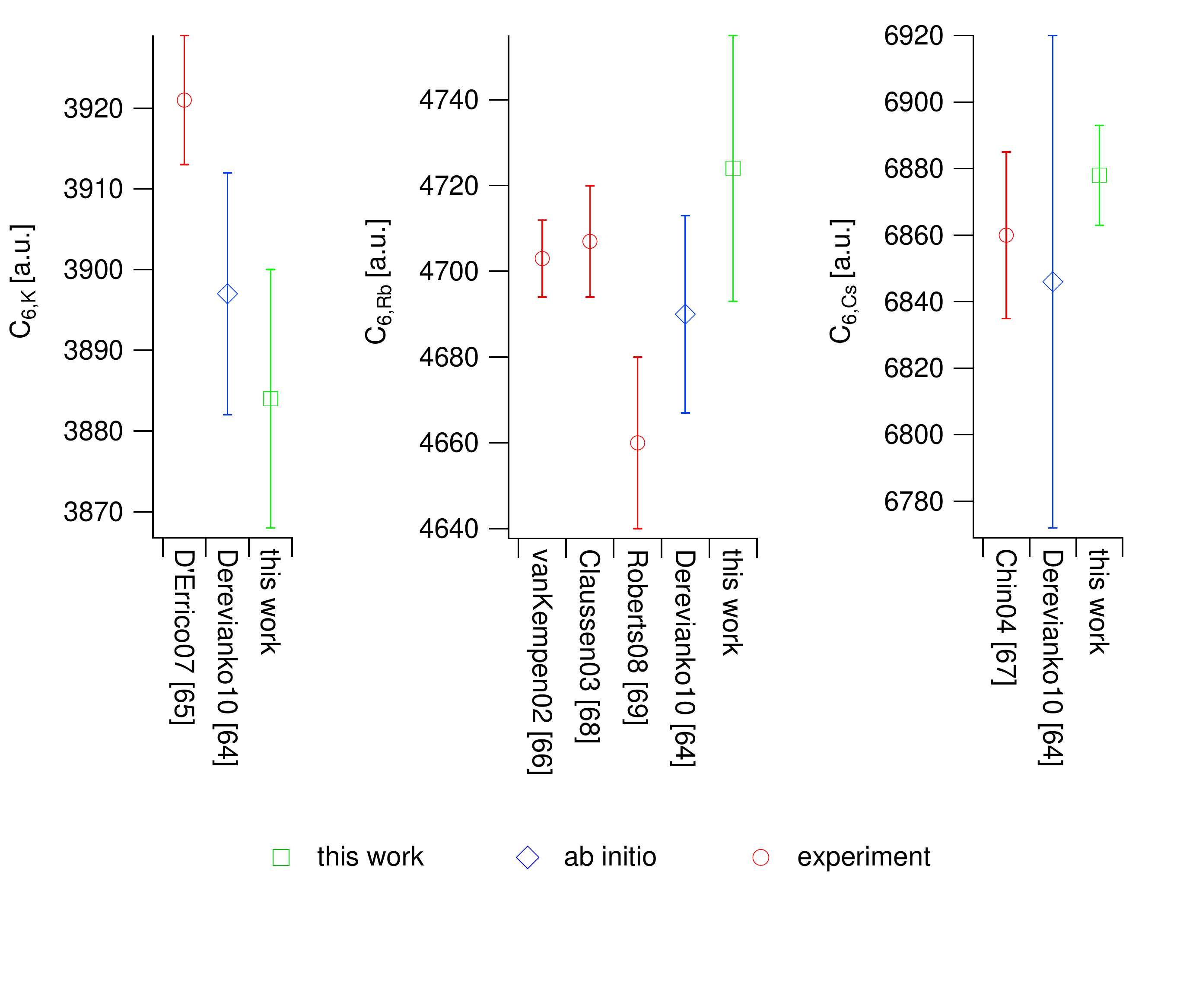}
\centering
\caption{\label{fig:c6}  Theoretical \cite{Derevianko2010} and experimental \cite{DErrico2007,VanKempen2002,Claussen2003,Roberts1998,Chin2004} $C_6$ values of K, Rb, and Cs from several different sources and the $C_6$ values determined in this article from polarizability measurements \cite{Gregoire2015}.
The experimental $C_6$ values were all determined from Feshbach resonance data. }
\end{figure}

Using \eqnref{eq:c6int_nop}, or equivalently Eqns.~(\ref{eq:c6int_pfactored}) and (\ref{eq:ar_tab}), our calculated $C_6$ values for Rb and Cs agree with recent theoretical and experimental $C_6$ values, as shown in \figref{fig:c6}. For K, our predicted $C_6$ is different from that measured by D'Errico \etal using Feshbach resonances by roughly 3$\sigma$.  Of course, this discrepancy may be at least partly explained by statistical errors in the $C_6$ and $\alpha(0)$ measurements for K atoms.   In the next section, however, we will explore how error in $\alpha_r(i\omega)$ used to construct $\alpha_{tab}(i\omega)$ for K could partly explain this discrepancy.

\begingroup
\begin{table}
\centering
\caption{\label{tab:c6} Homonuclear van der Waals $C_6$ coefficients, in atomic units, calculated using experimental static polarizabilities shown in Table \ref{tab:a0measurements} and tabulated dynamic polarizabilities from \cite{Derevianko2010}. The two contributions to the  uncertainty $\delta_{\alpha(0)}$ and $\delta_{\alpha_r(0)}$ for each $C_6$ value are, respectively, due to the uncertainties in measured $\alpha(0)$ and uncertainties estimated for Derevianko \etalnospace's values for $\alpha_r(i\omega)$ used to calculate $\alpha_{tab}(i\omega)$.   Derevianko \etal \cite{Derevianko1999} reported uncertainty in $\alpha_r(0)$ by using "an estimated 5\% error for the core polarizabilities, and a 10\% error for the remaining contributions to $\alpha_r(0)$."  Several other authors also estimate 5\% or 2\% error for $\alpha_{\mathrm{core}}$. }
\begin{center}
\begin{tabular}{llll}
\toprule
Atom & $C_6$ & $\delta_{\alpha(0)}$ & $\delta_{\alpha_r(0)}$  \\
\midrule
Na & 1558(11) & 10 & 1  \\
K  & 3884(16) & 7 & 14  \\ 
Rb & 4724(31) & 10 & 130  \\
Cs & 6879(15) & 13 & 7  \\
\bottomrule
\end{tabular}
\end{center}
\end{table}
\endgroup

To interpret the $C_6$ values that we report in Table \ref{tab:c6}, we compare these semi-empirical results to direct measurements and earlier predictions of $C_6$ in \figref{fig:c6}.  One sees that the uncertainty of $C_6$ measurements that we report based on atom interferometry measurements of polarizability are comparable to direct measurements \cite{DErrico2007,VanKempen2002,Chin2004} and slightly more precise than previous semi-empirical predictions \cite{Derevianko2010} .

\section{Determining residual polarizabilities empirically  \label{sec:testing_ar}}

\subsection{Using combinations of $\alpha(0)$ and $\tau$ measurements to report  $\alpha_r$ values} \label{ssec:tautoalphar}

While in Section  \ref{sec:pol2other} we demonstrated how to report atomic lifetimes from polarizability measurements and theoretical values for $\alpha_r(0)$, here we invert this procedure and use combinations of $\alpha(0)$, $\tau_{1/2}$, and $\tau_{3/2}$ measurements to place constraints on $\alpha_r(0)$.   For this, we solve \eqnref{eq:a0A} for $\alpha_r(0)$:
\begin{align}
	\alpha_r(0) = \alpha(0) - 2\pi\epsilon_0 c^3
	\left[  \frac{\tau_{1/2}^{-1}}{\omega_{D1}^4} + 2 \frac{\tau_{3/2}^{-1}}{\omega_{D2}^4} \right]
	\label{eq:arA}
\end{align}

\begin{figure}
\includegraphics[width=\linewidth,keepaspectratio]{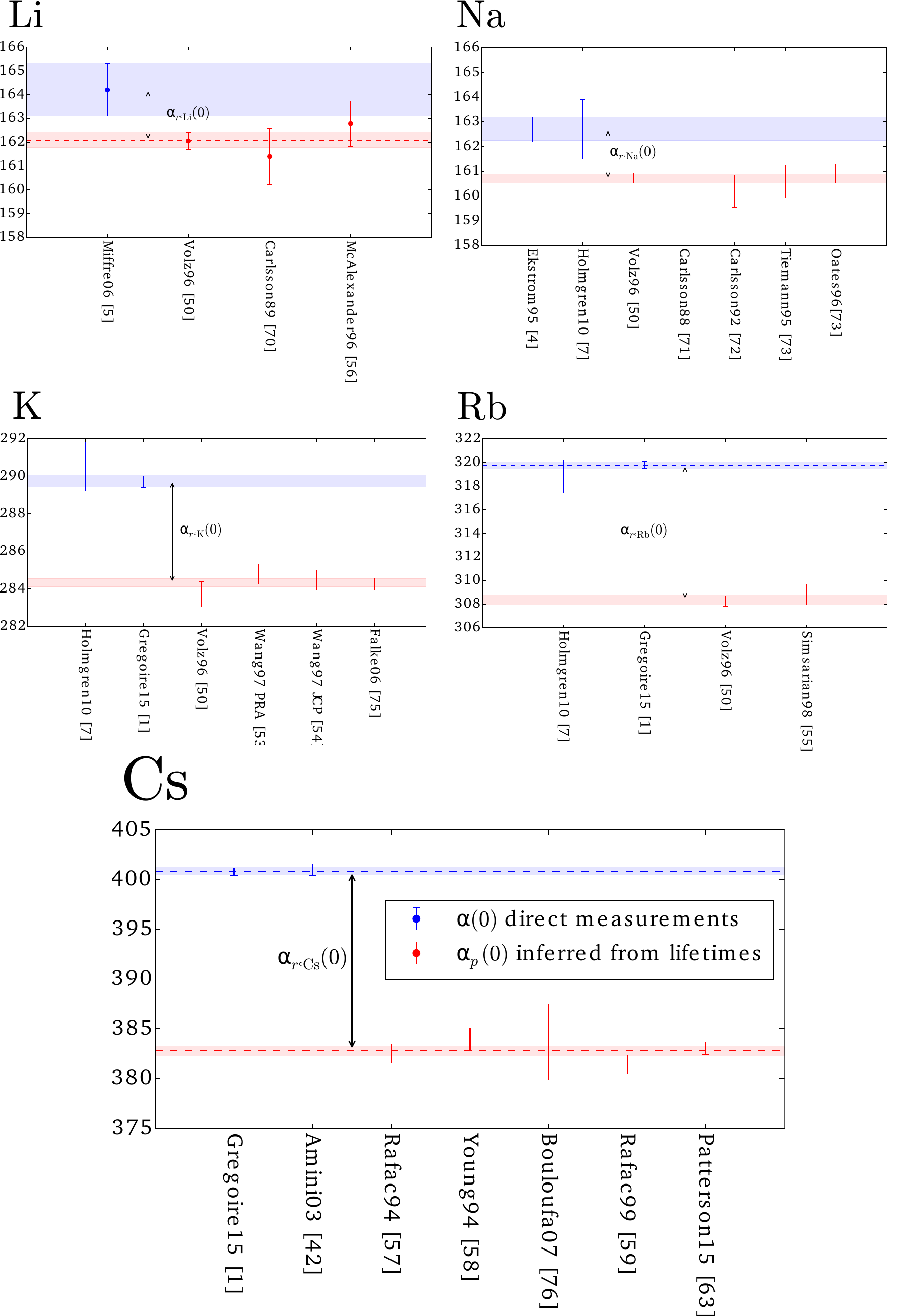}
\centering
\caption{\label{fig:ar_Cs} 
Differences between recent $\alpha(0)$ measurements \cite{Miffre2006,Ekstrom1995,Holmgren2010,Gregoire2015,Amini2003} and an average of $\alpha_{p,\mathrm{Cs}}$ values 
calculated from lifetime measurements
\cite{Volz1996,Carlsson1989,McAlexander1996,Carlsson1988,Carlsson1992,
Tiemann1996,Oates1996,Wang1997,Wang1997a,Falke2006a,Simsarian1998,
Rafac1994,Young1994,Bouloufa2007,Rafac1999,Patterson2015}
and $R$ values \cite{Johnson2008, Volz1996, Trubko2016, Leonard2015, Rafac1998}.
The uncertainties on those averages are used to calculate the resulting uncertainties in $\alpha_r(0)$.
}
\end{figure}

\figref{fig:ar_Cs} shows the difference between polarizability measurements $\alpha(0)$ and the inferred contribution to polarizability from the principal transitions $\alpha_p(0)$ based on lifetime measurements.   We take the weighted average of $\alpha_p(0)$ based on a collection of available lifetime measurements, and we use the weighted average of the two high-precision $\acs$ measurements.   
We obtain $\arli$ = 2(1), $\arna$ = 2.0(5), $\ark$ = 5.4(4), $\arrb$ = 11.4(5), and $\arcs$ = 18.1(5).
This analysis shows significantly nonzero $\alpha_r(0)$ values for Li, Na, K, Rb, and Cs based entirely on experimental data.

For Cs, this approach is sufficiently precise to empirically measure $\alpha_r(0)$ with 3\% uncertainty, which is similar to the uncertainty of theoretical values \cite{Safronova2006,Derevianko2002}.  The width of the blue and red bands in \figref{fig:ar_Cs} indicate the contributions to this uncertainty from the atom interferometry polarizability measurements and the uncertainty contributions from lifetime measurements.   In order to improve the accuracy of $\alpha(0)$ reported this way one would require improvements in both the polarizability measurements and the lifetime measurements.   

We can use a similar approach by combining polarizability measurements with \emph{ab initio} $|D_{ik}|$ calculations.  One of the highest-accuracy calculations of $|D_{D1}|$ was reported for Cs by Porsev \etal  \cite{Porsev2010}  in order to help interpret atomic parity violation experiments.  This $|D_{D1}|$ can be combined with $R$ \cite{Rafac1998} using equations (\ref{eq:a0D}) and (\ref{eq:Rdef}) as 
\begin{align}
	\alpha_r(0) = \alpha(0) - \frac{|D_{D1}|^2}{3\hbar} 
	\left[  \frac{1}{\omega_{D1}} + \frac{R}{\omega_{D2}} \right]
	\label{eq:arD}
\end{align}
This approach produces a somewhat lower value of $\alpha_r(0)$ = 16.5(4) with about 2.5\% uncertainty.    We compare the results using lifetimes and this result using the ratio of line strengths ($R$) and a calculated dipole matrix element with other results in \figref{fig:a_r_comp} at the end of this section.

\subsection{Using combinations of $\alpha(0)$ and $C_6$ measurements to report  $\alpha_r$ values} \label{ssec:C6toalphar}

\begin{figure}
\includegraphics[width=\linewidth,keepaspectratio]{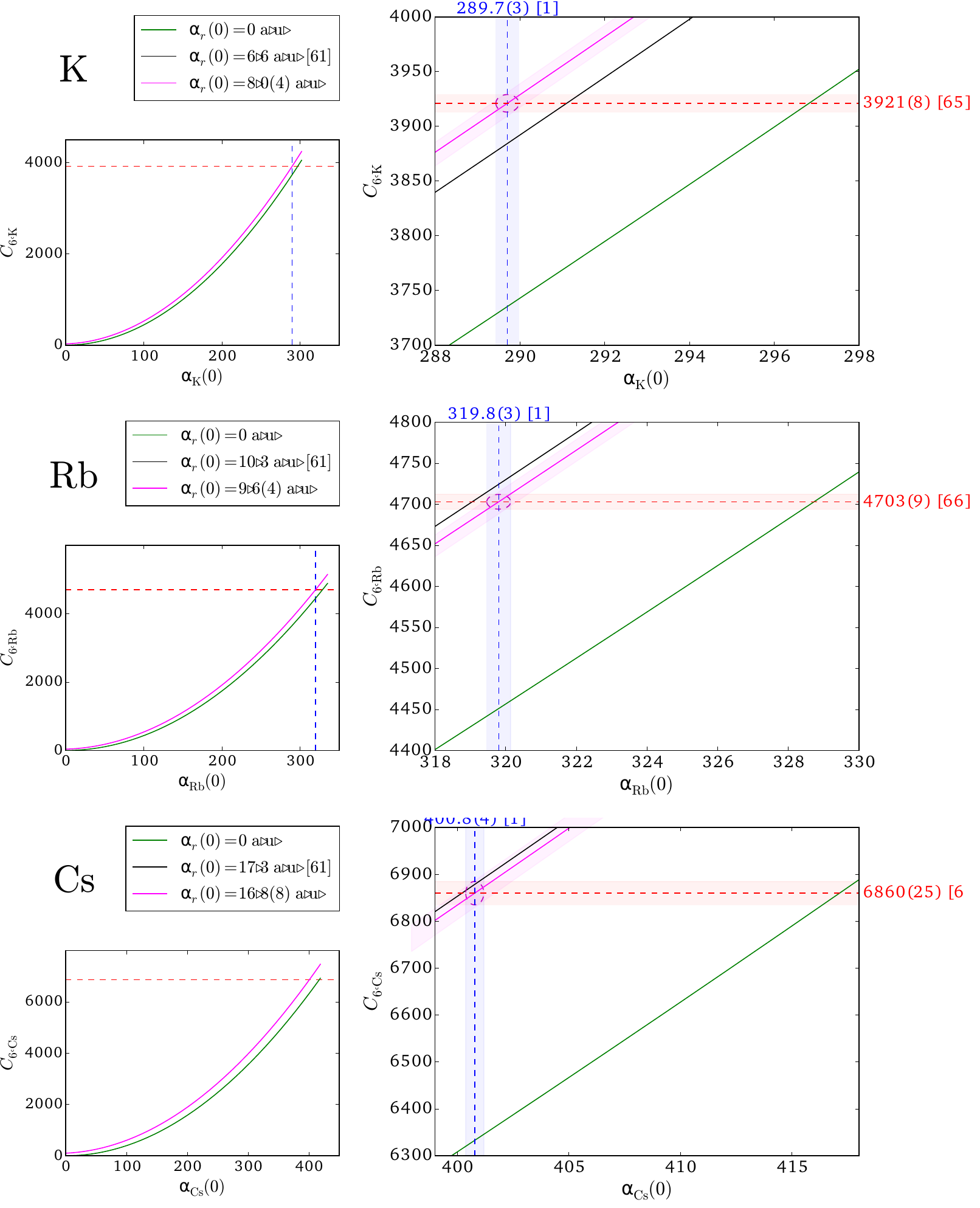}
\centering
\caption{\label{fig:alpha_c6_cs} 
$C_6$ versus $\alpha(0)$ for small deviations about experimental values of $\alpha(0)$ \cite{Gregoire2015}. The different curves correspond to different values of $\alpha_r(0)$: The pink curve corresponds to the value of $\alpha_r(0)$ inferred from experimental measurements of $C_6$ and $\alpha(0)$ \cite{Gregoire2015}, and the error bands on that curve represent the resulting uncertainty in $\alpha_r(0)$ due to uncertainty in $C_6$ and $\alpha(0)$. The black curve corresponds to the values of $\alpha(i\omega)$ tabulated by Derevianko \etal \cite{Derevianko2010}. 
Finally, the green line corresponds to $\alpha_r(0) = 0$, and the inset on each plot shows $C_6$ versus $\alpha(0)$ for a wider range of $\alpha(0)$. For these plots, we used the
$\ck$ measurement by D'Errico \etal \cite{DErrico2007}, $\crb$ by van Kempen \etal \cite{VanKempen2002}, and $\ccs$ by Chin \etal \cite{Chin2004}. In the smaller plots, we can see that $C_6$ versus $\alpha(0)$ is approximately quadratic.
}
\end{figure}

Earlier in Section \ref{sec:derivingC6} we demonstrated how to calculate van der Waals $C_6$ coefficients from polarizability measurements and assumptions about residual polarizabilities.   We can also invert this procedure, and analyze combinations of $\alpha(0)$ and $C_6$ measurements in order to place constraints on $\alpha_r(0)$.   For this we will assume the spectral function $\alpha_r(i\omega) / \alpha_r(0)$ is sufficiently known and simply factor out an overall scale factor for the static residual polarizability from the formula for $C_6$ \eqnref{eq:c6int_pfactored} as follows:
\begin{align}
	C_6 = \frac{3\hbar}{\pi} \int_0^\infty 
	\left[ 
		(\alpha(0)-\alpha_r(0))\frac{\alpha_p(i \omega)}{\alpha_p(0)} + 
		\alpha_r(0)\frac{\tilde{\alpha}_{r}(i \omega)}{\tilde{\alpha}_r(0)} 
	\right]^2 
	d\omega
	\label{eq:c6int_bothfactored}
\end{align}
where $\tilde{\alpha}_{r}(i \omega)$ and $\tilde{\alpha}_r(0)$ refer to values we infer from \eqnref{eq:ar_tab} using values tabulated by Derevianko \etal   We then plot predictions for $C_6$ versus predictions for $\alpha(0)$ parametric in hypothetical $\alpha_p(0)$ for different values of $\alpha_r(0)$. This is shown in \figref{fig:alpha_c6_cs} along with measurements of $C_6$ (red) and $\alpha(0)$ (blue).  Even on the graph with a large domain (small plots in Fig.~\ref{fig:alpha_c6_cs}) where one sees the generally quadratic dependence of $C_6$ on $\alpha(0)$,  it is evident that a model with $\alpha_r(0)$=0 is incompatible with the data.   On the expanded region of interest (larger plots in Fig.~\ref{fig:alpha_c6_cs}), one sees the intersection of $C_6$ and $\alpha(0)$ measurements specifies a value of $\alpha_r(0)$. For Cs, we obtain an $\alpha_r(0)$ = 16.8(8) that is consistent with $\alpha_r(0)$ found from the other two methods we have presented so far in Section \ref{sec:testing_ar}. This method is valuable because it relies on independent measurements of $C_6$ and $\alpha(0)$ to provide an empirical measurement of the size of $\alpha_r$.

These plots show the values of $\alpha_r(0)$ and the corresponding uncertainties that we would infer using experimental values (and their uncertainties) of $\alpha(0)$ and $C_6$.   
From these studies we find a best fit $\alpha_r(0)$ of 8.0(4) for K, 9.6(4) for Rb, and 16.8(8) for Cs.
The analysis for K highlights how the discrepancy between D'Errico \etalnospace's \cite{DErrico2007} $\ck$ measurement and the $\ck$ that we infer from our $\alpha(0)$ measurement \cite{Gregoire2015} could be explained in part by error in assumed $\ak$.

\begin{figure}
\includegraphics[width=\linewidth,keepaspectratio,valign=t]{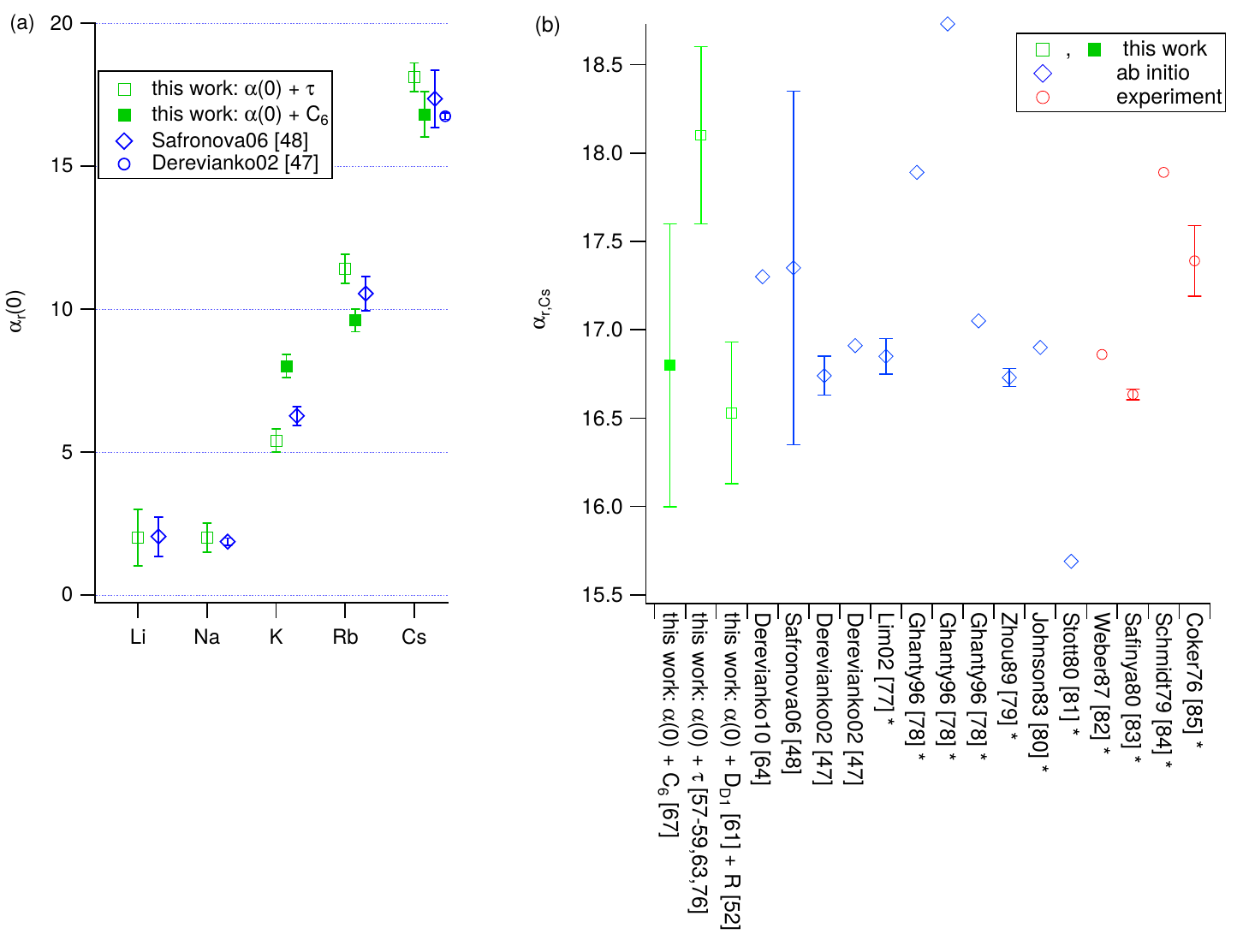}
\centering
\caption{\label{fig:a_r_comp} 
(a): $\alpha_r(0)$ values deduced by combining $\alpha(0)$ measurements \cite{Miffre2006,Ekstrom1995,Holmgren2010,Gregoire2015,Amini2003} with either $C_6$ measurements \cite{DErrico2007,VanKempen2002,Chin2004} or principal transition lifetime measurements \cite{Volz1996,Carlsson1989,McAlexander1996,Carlsson1988,Carlsson1992,
Tiemann1996,Oates1996,Wang1997,Wang1997a,Falke2006a,Simsarian1998,
Rafac1994,Young1994,Bouloufa2007,Rafac1999,Patterson2015} and $R$ values \cite{Johnson2008, Volz1996, Trubko2016, Leonard2015, Rafac1998}. These inferred values are compared to the theoretical $\alpha_r(0)$ values we used elsewhere in this work by Safronova \etal \cite{Safronova2006} and Derevianko and Porsev \cite{Derevianko2002}.
(b): $\arcs$ values deduced by combining measured $\acs$  \cite{Gregoire2015} with either measured $\ccs$ \cite{Chin2004}, principal transition lifetime \cite{Rafac1994,Young1994,Bouloufa2007,Rafac1999,Patterson2015} and $R_{\mathrm{Cs}}$ \cite{Rafac1998} measurements, or $\alpha_{p,\mathrm{Cs}}(0)$ inferred from calculated $|D_{\mathrm{D1,Cs}}|$ \cite{Porsev2010} and measured $R_{\mathrm{Cs}}$ \cite{Rafac1998}.
These inferred values are compared to  several theoretical calculations 
\cite{Derevianko2010,Safronova2006,Derevianko2002,Lim2002,Ghanty1996,Zhou1989,Johnson1983,Stott1980} and Cs ion polarizability measurements \cite{Weber1987,Safinya1980,Schmidt1979,Coker1976}.
The asterisk (*) indicates that the indicated references provided $\alpha_{\mathrm{core}}$ values which we converted to $\alpha_r(0)$ values by adding $\alpha_{v'} + \alpha_{cv} = 1.81 - 0.72 = 1.09$ (in atomic units). 
}
\end{figure}

\begingroup
\begin{table}
\centering
\caption{$\alpha_r(0)$ values deduced by combining $\alpha(0)$ measurements \cite{Miffre2006,Ekstrom1995,Holmgren2010,Gregoire2015,Amini2003} with either $C_6$ measurements \cite{DErrico2007,VanKempen2002,Chin2004}, principal transition lifetime measurements \cite{Volz1996,Carlsson1989,McAlexander1996,Carlsson1988,Carlsson1992,
Tiemann1996,Oates1996,Wang1997,Wang1997a,Falke2006a,Simsarian1998,
Rafac1994,Young1994,Bouloufa2007,Rafac1999,Patterson2015} and $R$ values \cite{Johnson2008, Volz1996, Trubko2016, Leonard2015, Rafac1998}, or $|D_{D1}|$ \cite{Porsev2010} and $R$ values \cite{Rafac1998}. These inferred values are compared to the theoretical $\alpha_r(0)$ values we used elsewhere in this work by Safronova \etal \cite{Safronova2006} and Derevianko and Porsev \cite{Derevianko2002}.
}
\label{tab:a_r_comp}
\begin{tabular}{llllll}
\toprule
atom &  $\alpha(0)+\tau$ [$+R$] & $\alpha(0)+C_6$ &  $\alpha(0)+|D_{D1}|+R$ & \abinitio &     \\
\midrule
Li & 2(1) &  & & 2.04(69) \cite{Safronova2006} \\
Na & 2.0(5) &  & & 1.86(12) \cite{Safronova2006} \\
K  & 5.4(4) & 8.0(4) & & 6.23(33) \cite{Safronova2006} \\
Rb & 11.4(5) & 9.6(4) & & 10.54(60) \cite{Safronova2006} \\
Cs & 18.1(5) & 16.8(8) & 16.5(4) & 17.35(100) \cite{Safronova2006} & 16.74(11) \cite{Derevianko2002} \\ 
\arrayrulecolor{black} \bottomrule
\end{tabular}
\end{table}
\endgroup

\figref{fig:a_r_comp}a shows the $\arcs$ values we inferred from lifetime measurements, van der Waals $C_6$ coefficients, and Porsev \etalnospace's calculated $|D_{D1,\mathrm{Cs}}|$. Our results are compared to \abinitio calculations of $\arcs$ \cite{Derevianko2010,Safronova2006,Derevianko2002} as well as \abinitio calculations of $\alpha_{\mathrm{core,Cs}}$ \cite{Lim2002,Ghanty1996,Zhou1989,Johnson1983,Stott1980} to which we added $\alpha_{v'} + \alpha_{cv}$. Also among the comparisons in \figref{fig:a_r_comp}a are measurements of Cs$^+$ ionic polarizability \cite{Weber1987,Safinya1980,Schmidt1979,Coker1976}, which approximates $\alpha_{\mathrm{core,Cs}}$, again adjusted by adding $\alpha_{v'} + \alpha_{cv}$. \figref{fig:a_r_comp}b shows our results alongside the \abinitio $\alpha_r(0)$ values calculated by Safronova \etal \cite{Safronova2006} and Derevianko and Porsev \cite{Derevianko2002} for Li, Na, K, Rb, and Cs.

\figref{fig:a_r_comp} shows some disagreement between our $\alpha(0) + \tau$ and $\alpha(0) + C_6$ methods, especially with regard to K and Rb. There are several possible contributors to such disagreement. While the $\alpha(0) + \tau$ results were based on an average of several, independently-measured lifetimes, both of our methods relied on only one (or, in the case of Cs, two) $\alpha(0)$ measurements and our $\alpha(0) + C_6$ method relied on a single $C_6$ measurement. Therefore, statistical variation or systematic errors that were not accounted for in those $\alpha(0)$ or $C_6$ measurements could have a significant effect on our reported $\alpha_r(0)$. Also, it is important to note that our $\alpha(0) + C_6$ method relied on a single set of $\alpha(i\omega)$ calculated using one specific theoretical approach \cite{Derevianko2010}, and that there are other theoretical approaches that could lead to different values of $\alpha(i\omega)$. 

The uncertainties in \abinitio $\alpha_r(0)$ predictions by Safronova \etal \cite{Safronova2006} and Derevianko and Porsev \cite{Derevianko2002} are comparable to or smaller than the uncertainties on our fully-empirical ($\alpha(0) + \tau$) and semi-empirical ($\alpha(0) + C_6$) results. This fact, combined with the aforementioned possible contributors to disagreement between our results, suggests that \abinitio methods are still the prefered way of obtaining $\alpha_r(0)$ values for use in other analyses. Even so, it is valuable to develop the methods of analysis demonstrated in this paper so that when more accurate $\alpha(0)$, $\tau$, and $C_6$ measurements become available, then $\alpha_r(0)$ can be determined with higher accuracy using these methods.


The theoretical $\alpha_r(0)$ predictions by Safronova et al. \cite{Safronova2006} have an
uncertainty of 6\%, which is just slightly larger than the 5\% or 3\%
uncertainties of the experimental $\alpha_r(0)$ determinations that we
reported for Cs in Sections \ref{ssec:tautoalphar} and \ref{ssec:C6toalphar}.    However, we acknowledge that
there is a 10\% deviation between the all-experimental result for $\arcs$
reported in Section \ref{ssec:tautoalphar} using $\alpha(0)$ and $\tau$ measurements as
compared to the semi-empirical result for $\arcs$ that we reported in
Section \ref{ssec:C6toalphar} using $\alpha(0)$ and $C_6$ measurements combined with the
theoretical spectral function 
$\alpha_r(i\omega)/\alpha_r(0)$.
Furthermore, the
uncertainty in the theoretical $\arcs$ prediction by Derevianko
and Porsev \cite{Derevianko2002} is significantly smaller, approximately 0.6\%  (and
this was partly verified with independent measurements of $\alpha_{\mathrm{core}}$
using Rydberg spectroscopy \cite{Zhou1989}).  So, it is possible that ab initio
methods are still the preferred way of obtaining $\alpha_r(0)$ values for use
in other analyses.   Even so, we conclude that it is valuable to
develop the methods of analysis demonstrated in this paper so that
when more accurate $\alpha(0)$, $\tau$, and $C_6$ measurements become available, then
$\alpha_r(0)$ can be determined with higher accuracy using these methods.   In
the future, combining measurements of $\alpha_{\mathrm{core}}$ from Rydberg spectroscopy
with higher accuracy measurements of $\alpha_r(0)$ could provide more direct constraints on $\alpha_{v'} + \alpha_{cv}$, and thus $\alpha_{\mathrm{tail}}$.

\section{Discussion} \label{sec:discussion}


In this paper we reported measurements of the static polarizabilities
of K, Rb, and Cs atoms with reduced uncertainties.    We made these
measurements with an atom interferometer and an electric field
gradient using data originally reported in [1].  We described in
Section \ref{sec:revPolUnc} how we reduced the systematic uncertainty in $\alpha(0)$ measurements from
0.15\% to 0.10\% by improving the calibration of the electric field.  To
our knowledge, these are now the most precise measurements of atomic
polarizabilities that have been made using any method for K, Rb, and Cs
atoms.   For Cs in particular, the improvement described in this paper
enabled us to report a value of $\acs$ with slightly smaller
uncertainty than Amini and Gould’s measurement of $\acs$ that they
obtained using an atomic fountain experiment \cite{Amini2003}.  Currently, this means that atom
interferometer experiments have made the most accurate measurements of
atomic polarizabilities for all of the alkali metal atoms Li, Na, K,
Rb, and Cs.

In Section \ref{sec:pol2other} we demonstrated how to analyze these measurements of
atomic polarizabilities in order to infer oscillator strengths,
lifetimes, transition matrix elements, line strengths, and van der
Waals $C_6$ coefficients for all the alkali metal atoms, as we did in
Tables \ref{tab:ff} and \ref{tab:c6}.   We referred to the idea chart in \figref{fig:ideachart} to review
how these quantities are interrelated, and we described this more
explicitly with Eqns.~\ref{eq:alphaf}-\ref{eq:Rdef} and \ref{eq:c6int}-\ref{eq:c6int_nop}.   Building on these
interrelationships, we specifically used measurements of static
polarizabilities obtained with atom interferometry, empirical ratios
of line strengths $R$ (some of which were also obtained with atom
interferometry), and theoretical values for residual polarizabilities
in order to deduce the lifetimes of excited $np_J$ states for all of
the alkali metal atoms with unprecedented accuracy.  These methods
also allow us to use static polarizability measurements as a
semi-empirical benchmark to test \abinitio predictions of principal
$ns$-$np_J$ transition matrix elements for alkali metal atoms.
Furthermore, we used these methods to test the extent to which
measurements of different atomic properties such as lifetimes,
branching ratios, line strengths, polarizabilities, and van der Waals
interactions agree with one another, as shown in Figs.~\ref{fig:tau_cs} and \ref{fig:c6}.

Then in Section \ref{sec:testing_ar} we explored new methods to infer residual
polarizability $\alpha_r(0)$ values by combining measurements of atomic
polarizabilities with independent measurements of lifetimes or $C_6$
coefficients. This constitutes a novel, all-experimental method to test
several theoretical $\alpha_r(0)$ predictions.   Using this approach it is
clear that atom interferometry measurements of atomic polarizabilities
are sufficiently precise to detect non-zero residual polarizabilities
for all of the alkali metal atoms, and can measure $\alpha_r(0)$ with as
little as 3\% uncertainty for Cs atoms.  This procedure also provides a
motivation for next generation $C_6$, $\alpha(0)$, $\tau$, and tune-out wavelength
measurements that can be combined with one another to more accurately
determine $\alpha_r(0)$ values that are needed in order to test atomic
structure calculations that are relevant for interpreting atomic
parity violation and atomic clocks.

\section{Acknowledgements}

This work is supported by NSF Grant No. 1306308 and a NIST PMG. M.D.G. and R.T. are grateful for NSF GRFP Grant No. DGE-1143953 for support. N.B. acknowledges support from the NASA Space Grant Consortium at the University of Arizona.

\appendix

\section{\label{sec:ar}}

\begin{table}
\centering
\caption{Contributions to residual polarizability $\alpha_r(0) = \alpha_{v'}(0) + \alpha_{\mathrm{core}}(0) + \alpha_{cv}(0)$  in atomic units.  The quantity $\alpha_{v'}(0)$ is the sum of all the contributions from the valence electron $ns$-$n'p_J$ transitions with $n'>n$ using \eqnref{eq:alphaf}.  This includes $\alpha_{\mathrm{tail}}(0)$.    Values in bold are used to produce the results presented  in Table \ref{tab:ff} are in bold.}
\label{tab:ar}
\begin{tabular}{llllllll}
\toprule
atom &  $\alpha_{v'}(0)$ &  $\alpha_{\mathrm{core}}(0)$ & &  $\alpha_{cv}(0)$ &  $\alpha_r(0)$  &   \\
\midrule
Li   &   &   0.189(9)   &  \cite{Johnson1983}  $^{(b)}$     &   &$\mathbf{2.04(69)} $ & \cite{Safronova2006} \\
Li   &   &   0.192   &  \cite{Johnson1996}   &   &  &  \\
\arrayrulecolor[gray]{0.8} \midrule
Na   & 0.81 $^{(a)}$ & 0.94(5)  &\cite{Johnson1983} $^{(b)}$     &               &   & \\
Na   &                & 1.00(4)        &\cite{Lim2002}      &            & $\mathbf{1.86(12)} $ & \cite{Safronova2006} \\                                   
\midrule
K    & 0.72 $^{(a)}$      & 5.46(27)  &\cite{Johnson1983} $^{(b)}$        &      &       & \\
K    & 0.90 \cite{Safronova2008} & 5.50  &     \cite{Safronova2008}   &    & &     \\
K    &                    & 5.52(4)  &  \cite{Lim2002}   &    &     & \\
K    &         & 5.50 &\cite{Safronova2013}       &  -0.18 \cite{Safronova2013}   & $\mathbf{6.26(33)}$  & \cite{Safronova2006}   \\
\midrule
Rb   & 1.32 $^{(a)}$ & 9.08(45) &\cite{Johnson1983} $^{(b)}$     &        &    &                   \\
Rb   &    & 9.11(4) & \cite{Lim2002}     &            &  10.70(22) &\cite{Leonard2015} $^{(c)}$  \\
Rb   &    & 9.11(4) &\cite{Safronova2003}   &  -0.30 \cite{Safronova2003}  &  $\mathbf{10.54(60)}$ &\cite{Safronova2006}  \\
\midrule
Cs  & 1.60 $^{(a)}$    & 15.8(8)&  \cite{Johnson1983} $^{(b)}$  &   -0.72 \cite{Derevianko2002} &  17.35(100)  & \cite{Safronova2006}  \\
Cs  &    & 15.8(1)& \cite{Lim2002}  &              &    16.91 &\cite{Derevianko2002} \\
Cs  & 1.81 \cite{Derevianko2002} & 15.81 &   \cite{Derevianko2002}   &   &  $\mathbf{16.74(11)}$ &\cite{Derevianko2002}   \\
Cs  &  & 16.3(2) &   \cite{Coker1976} $^{(d)}$   &   &    &   \\
Cs  &  &  15.17 &   \cite{rosseinsky1994} $^{(d)}$    &   &    &   \\  
Cs  &  & 15.54(3) &   \cite{Safinya1980}  $^{(e)}$ &   &    &   \\
Cs  &  & 15.82(3) &   \cite{ruff1980}  $^{(e)}$ &   &    &   \\
Cs  &  & 15.770(3) &   \cite{Weber1987} $^{(e)}$  &   &    &   \\
Cs  &  & 17.64 &   \cite{Ghanty1996} $^{(f)}$   &   &    &   \\
\arrayrulecolor{black} \bottomrule
\end{tabular}
\newline
$^{(a)}$ Calculated using $f_{ik}$ values from NIST \cite{NIST}  for $n-n'$ transitions with $n' = n+1$ to $n+5$.    \newline
$^{(b)}$ For $\alpha_{\mathrm{core}}$ from \cite{Johnson1983} we list a fractional uncertainty of 5\% as suggested in reference \cite{Safronova2003}. \newline
$^{(c)}$ Reference \cite{Leonard2015} calculated ($\alpha_{\mathrm{core}} + \alpha_{cv}$) = 8.71(9) and $\alpha_r=10.70(22)$ at $\omega = 2\pi c / 790$nm.   \newline
$^{(d)}$  from studies of ions in solid crystals \newline
$^{(e)}$  from Rydberg spectroscopy data \newline
$^{(f)}$  A result from DFT calculations \newline
\end{table}

\newpage
\bibliographystyle{mdpi}


\bibliography{library}


\end{document}